\documentclass[twocolumn]{aastex631}

\usepackage{booktabs}
\usepackage{xcolor}
\usepackage{amsmath,amssymb}
\usepackage{CJK}

\shorttitle{AASTeX v6.3.1 Sample article}
\shortauthors{Roberts et al.}
\begin{document}
\begin{CJK*}{UTF8}{gbsn}

\title{LoTSS Jellyfish Galaxies IV: Enhanced Star Formation on the Leading Half of Cluster Galaxies and Gas Compression in IC3949}

\author[0000-0002-0692-0911]{Ian D. Roberts}
\affiliation{Leiden Observatory, Leiden University, PO Box 9513, 2300 RA Leiden, The Netherlands}

\author{Maojin Lang (郎茂锦)}
\affiliation{Leiden Observatory, Leiden University, PO Box 9513, 2300 RA Leiden, The Netherlands}

\author{Daria Trotsenko}
\affiliation{Leiden Observatory, Leiden University, PO Box 9513, 2300 RA Leiden, The Netherlands}

\author[0000-0003-0618-8473]{Ashley Bemis}
\affiliation{Leiden Observatory, Leiden University, PO Box 9513, 2300 RA Leiden, The Netherlands}

\author[0000-0002-1768-1899]{Sara L. Ellison}
\affiliation{Department of Physics \& Astronomy, University of Victoria, Finnerty Road, Victoria, British Columbia, V8P 1A1, Canada}

\author[0000-0001-7218-7407]{Lihwai Lin}
\affiliation{Institute of Astronomy and Astrophysics, Academia Sinica, No. 1, Section 4, Roosevelt Road, Taipei 10617, Taiwan}

\author[0000-0002-1370-6964]{Hsi-An Pan}
\affiliation{Department of Physics, Tamkang University, Tamsui Dist., New Taipei City 251301, Taiwan}

\author[0000-0003-1581-0092]{Alessandro Ignesti}
\affiliation{INAF- Osservatorio astronomico di Padova, Vicolo Osservatorio 5, IT-35122 Padova, Italy}

\author[0000-0002-4826-8642]{Sarah Leslie}
\affiliation{Leiden Observatory, Leiden University, PO Box 9513, 2300 RA Leiden, The Netherlands}

\author[0000-0002-0587-1660]{Reinout J. van Weeren}
\affiliation{Leiden Observatory, Leiden University, PO Box 9513, 2300 RA Leiden, The Netherlands}



\begin{abstract}

\noindent With MaNGA integral field spectroscopy, we present a resolved analysis of star formation for 29 jellyfish galaxies in nearby clusters, identified from radio continuum imaging taken by the Low Frequency Array. Simulations predict enhanced star formation on the ``leading half" of galaxies undergoing ram pressure stripping, and in this work we report observational evidence for this elevated star formation. The dividing line (through the galaxy center) that maximizes this star formation enhancement is systematically tied to the observed direction of the ram pressure stripped tail, suggesting a physical connection between ram pressure and this star formation enhancement. We also present a case study on the distribution of molecular gas in one jellyfish galaxy from our sample, IC3949, using ALMA CO J=1-0, HCN J=1-0, and HCO$^+$ J=1-0 observations from the ALMaQUEST survey. The $\mathrm{H_2}$ depletion time (as traced by CO) in IC3949 ranges from $\sim\!1\,\mathrm{Gyr}$ in the outskirts of the molecular gas disk to $\sim\!11\,\mathrm{Gyr}$ near the galaxy center. IC3949 shows a clear region of enhanced star formation on the leading half of the galaxy where the average depletion time is $\sim\!2.7\,\mathrm{Gyr}$, in line with the median value for the galaxy on the whole. Dense gas tracers, HCN and HCO$^+$, are only detected at the galaxy center and on the leading half of IC3949. Our results favour a scenario in which ram pressure compresses the interstellar medium, promoting the formation of molecular gas that in turn fuels a localized increase of star formation.

\end{abstract}



\section{Introduction} \label{sec:intro}

Galaxies in clusters are subject to a host of physical processes not experienced by galaxies isolated in the field. As galaxies infall toward the minimum of the cluster potential they are perturbed through gravitational interactions with other galaxies as well as hydrodynamical interactions with the dense intracluster medium (ICM). The perturbations can affect both the stellar distribution within galaxies as well as gas in the interstellar medium (ISM). This is seen through observations at low-redshift showing that the ratio between the number of quiescent and star-forming (or roughly speaking, red and blue) galaxies is strongly enhanced in galaxy clusters compared to galaxies in lower density environments or the field \citep[e.g.][]{dressler1980,postman1984,croton2005,peng2010,wetzel2012,lin2014,brown2017,jian2018,davies2019}. This is interpreted as evidence for expedited quenching of star formation in dense environments, likely a result of the cluster-specific processes that are outlined below.
\par
Gravitational interactions include tidal effects that can remove both gas and stars from galaxies \citep[e.g.][]{mayer2006,chung2007} as well as repeated impulsive interactions between galaxies (``harassment'', e.g.\ \citealt{moore1996}). Excluding the central brightest cluster galaxy, mergers are less common in galaxy clusters than lower density environments (i.e.\ groups) due to the high relative velocities between cluster galaxies, but still occur at a reduced rate \citep[e.g.][]{jian2012}. Hydrodynamical interactions with the hot ICM can both prevent the accretion of fresh gas onto galaxies (``starvation'', e.g.\ \citealt{larson1980,peng2015}) and directly remove warm/cold gas from galactic disks through ram pressure stripping (RPS, e.g.\ \citealt{gunn1972,quilis2000}). In practice the prevention of gas accretion via starvation is likely driven by a modest ram pressure that is able to remove the weakly bound circumgalactic medium from galaxies, thus while RPS and starvation are often treated as separate processes, they are in all likelihood closely connected to one another. Viscous stripping and thermal evaporation of cold gas also likely play some role in gas removal from cluster galaxies \citep[e.g.][]{cowie1977,nulsen1982}. While it is difficult to reliably disentangle the effects of these various hydrodynamical mechanisms observationally, it is clear that they have important implications for the future of gas content and star formation in cluster galaxies \citep[][and references therein]{cortese2021,boselli2021_review}.
\par
While the importance of different quenching mechanisms is still a matter of debate, some consensus has emerged that RPS plays a significant role \citep[e.g.][]{gavazzi2001,yagi2010,poggianti2017,boselli2018,jaffe2018,maier2019,roberts2019,ciocan2020,boselli2021_review,cortese2021}. This is in part due to the identification of so-called ``jellyfish galaxies'' in nearby groups and clusters, which has allowed RPS to be studied more directly. These are star-forming galaxies observed to have one-sided tails extending beyond the galaxy disk that are believed to be the product of RPS. These tails have been observed across the electromagnetic spectrum, including ionized gas traced by X-rays \citep[e.g.][]{sun2006,sun2010,poggianti2019b,sun2021}, the UV continuum \citep[e.g.][]{smith2010,boissier2012,george2018,mahajan2022}, and $\mathrm{H\alpha}$ emission \citep[e.g.][]{gavazzi2001,yagi2010,poggianti2017,boselli2018}; atomic gas traced by hydrogen $21\,\mathrm{cm}$ emission \citep[e.g.][]{kenney2004,oosterloo2005,chung2009,kenney2015,hess2022}; molecular gas traced by rotational transitions of the carbon monoxide molecule \citep{jachym2017,lee2018,jachym2019,moretti2020,moretti2020b}; and cosmic ray electrons traced by synchrotron emission in the radio continuum \citep[e.g.][]{gavazzi1987,murphy2009,chen2020,roberts2021_LOFARclust,roberts2021_LOFARgrp}. The strength of RPS scales with $\sim \rho v^2$, where $\rho$ is the density of the ICM and $v$ is the relative velocity between the galaxy and the ICM. Given the the dependence on ICM density and velocity dispersion, galaxies should experience stronger ram pressure in high-mass clusters compared to low-mass groups; a trend which is consistent with results from both simulations and observations \citep[e.g.][]{yun2019,roberts2021_LOFARgrp,roberts2022_UNIONS} and is also consistent with the fact that galaxy quenched fractions in groups and clusters monotonically increase with host halo mass \citep[e.g.][]{kimm2009,wetzel2012}. If the strength of ram pressure exceeds the restoring gravitational potential of the galaxy then gas stripping can occur, which if efficient, will remove a galaxy's reservoir of cold gas and thus quench star formation. If RPS is able to directly strip molecular gas then this quenching will occur very rapidly ($< 1\,\mathrm{Gyr}$); however, if only atomic gas is efficiently removed then the quenching timescale will be longer and set by the depletion time of the remaining molecular gas reserve.
\par
There is increasing evidence that the RPS does not cause a simple, monotonic decrease in the star formation rate (SFR), and instead ram pressure can temporarily enhance star formation in galaxies \citep[e.g.][]{gavazzi2001,vulcani2018_sf,roberts2020,boselli2021_ic3476,cramer2021,durret2021}. This is likely a result of ISM compression induced by ram pressure that leads to high gas densities, efficient formation of molecular gas, and strong star formation \citep[e.g.][]{schulz2001,moretti2020,moretti2020b,cramer2021}. Theoretically this SFR enhancement should be strongest on the ``leading half'' of the galaxy where the ISM will be highly compressed by ram pressure (i.e.\ opposite to the direction of the stripped tail). From the EAGLE simulations \citep{schaye2015}, there is evidence that star formation is systematically enhanced by a factor of $\sim\!1.1$ to $\sim\!1.5$ on the leading half of simulated galaxies in groups and clusters \citep{troncoso-iribarren2020}. \citeauthor{troncoso-iribarren2020} use the direction of the three dimensional velocity vector in order to split each galaxy into a leading and trailing half. It is more difficult to directly test this prediction observationally as full velocity information is not observable and projection effects are unavoidable, but there are isolated examples of observed galaxies with asymmetric star formation that is thought to be induced by ram pressure on the leading half \citep[e.g.][]{gavazzi2001,lee2016,boselli2021_ic3476,roberts2022_perseus,hess2022}.
\par
In this work we observationally address the question of whether or not star formation is enhanced on the leading half of galaxies undergoing RPS. We use a sample of 29 jellyfish galaxies \citep{roberts2021_LOFARclust} identified on the basis of $144\,\mathrm{MHz}$ radio continuum tails observed by the LOw Frequency ARray (LOFAR, \citealt{vanhaarlem2013}) that also have public integral field spectroscopy (IFS) from the Mapping Nearby Galaxies at APO (MaNGA, \citealt{bundy2015,abdurrouf2022}) survey. Based on the observed tail directions relative to their host cluster centers \citep{roberts2021_LOFARclust}, as well as the rarity of close galaxy neighbours for this sample of jellyfish galaxies, we believe that these tails are primarily produced via RPS though we cannot fully rule out that tidal effects may contribute to some of the radio continuum asymmetries. This low frequency continuum flux is tracing synchrotron emission from cosmic ray electrons moving through magnetic fields that extend from the stellar disk along the stripped tail.  The LOFAR observations of these stripped tails are critical for this analysis as they allow us to estimate the direction of motion across the plane of the sky (i.e.\ opposite to the direction of the LOFAR radio continuum tail, modulo projection), thus defining the leading half and trailing half of each galaxy. The MaNGA observations are equally important as they allow us to resolve the spatial distribution of star formation across these galaxies with extinction corrected $\mathrm{H\alpha}$ maps. This unique dataset is ideal for observationally testing the prediction of enhanced star formation on the leading half of jellyfish galaxies. Furthermore, this work introduces a sample of LOFAR-MaNGA jellyfish galaxies that can be utilized moving forward to study the resolved properties across the disks of galaxies undergoing RPS in nearby clusters. Finally, we present a case study on the connection between star formation and molecular gas in the jellyfish galaxy IC3949, including a region of ram pressure induced star formation. This analysis adds to the rapidly growing number of studies that probe the molecular gas content of galaxies undergoing RPS \citep[e.g.][]{vollmer2012,jachym2014,lee2017,jachym2017,jachym2019,moretti2020,moretti2020b,cramer2020,cramer2021,cramer2022,lee2022,zabel2022,villanueva2022}, and to our best knowledge this is the first analysis of dense molecular gas tracers (HCN and HCO$^+$) in a jellyfish galaxy.
\par
The outline of this paper is as follows. In Section~\ref{sec:data} we describe the galaxy samples used in this work along with the derived data products. In Section~\ref{sec:enhanced_sfr} we present evidence for enhanced star formation on the leading half of jellyfish galaxies. In Section~\ref{sec:ic3949} we present a case-study on the connection between star formation and molecular gas in the jellyfish galaxy IC3949. Finally, in Section~\ref{sec:discussion_conclusions} we provide a brief discussion of the main results of this work and give concluding statements. Throughout this manuscript we assume a flat Lambda cold dark matter cosmology with $(\Omega_M, \Omega_\Lambda, H_0) = (0.3, 0.7, 70\,\mathrm{km\,s^{-1}\,Mpc^{-1}})$ and a Salpeter \citep{salpeter1955} initial mass function.

\section{Data} \label{sec:data}

In this work we use public IFS from the MaNGA survey to explore the spatial distribution of star formation in both jellyfish galaxies in nearby clusters and a matched control sample of non-cluster galaxies. Below we describe the selection of these two galaxy samples. For both the control and jellyfish samples we use integrated galaxy stellar masses from the Pipe3D catalogue \citep{sanchez2016a,sanchez2016b} and redshifts from SDSS spectroscopy.

\subsection{Jellyfish Galaxy Sample} \label{sec:data_jellyfish}

\begin{figure*}
    \centering
    \includegraphics[width=\textwidth]{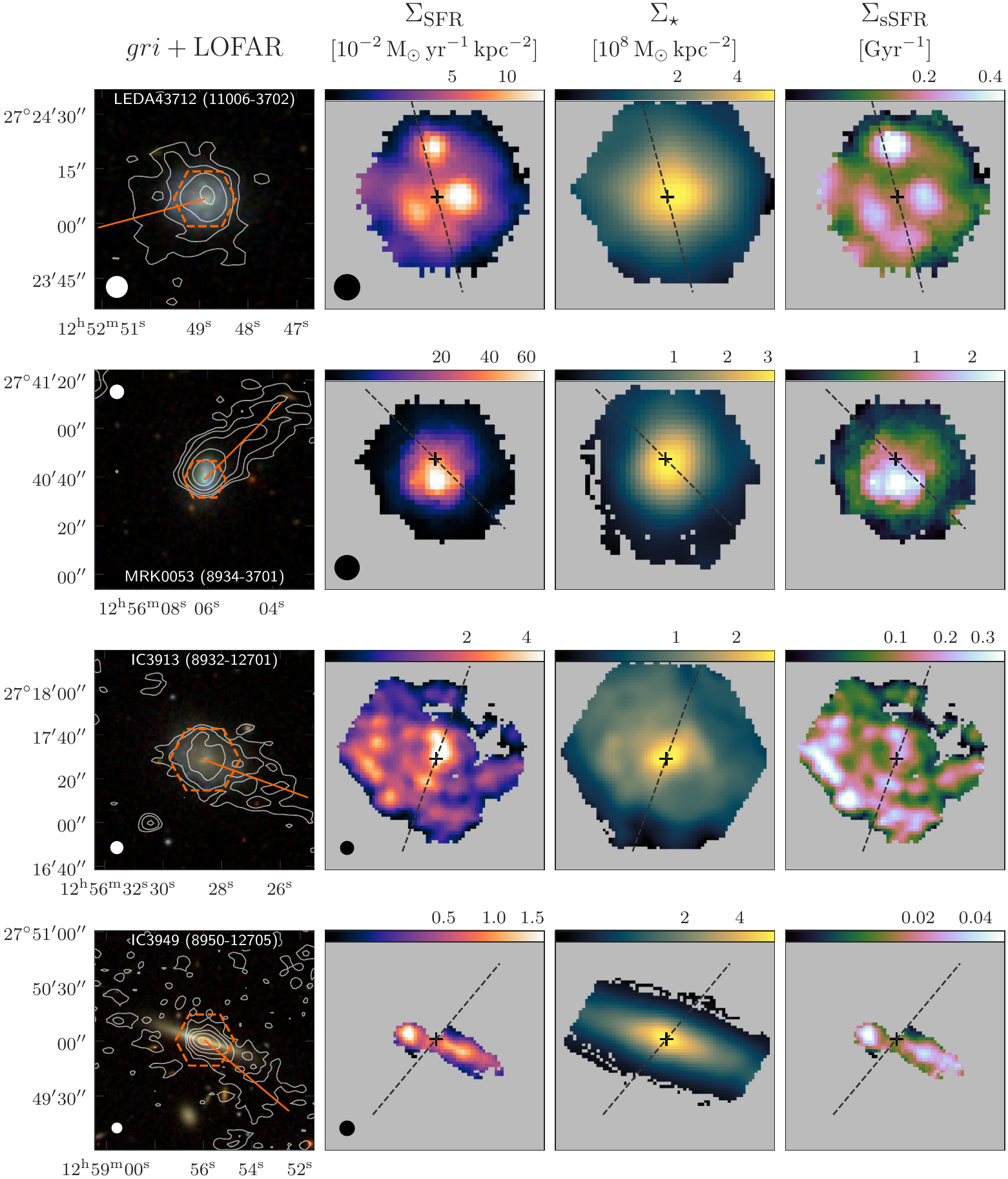}
    \caption{\emph{Left to right:} SDSS optical image plus LOFAR contours at the $(2, 4, 8, 16, ...)\,\sigma$ levels, SFR surface density map, stellar mass surface density map, and sSFR surface density map. The white circle shows the $6''$ LOFAR beam, the orange hexagon and line show the MaNGA field-of-view and the tail orientation, the black circle shows MaNGA FWHM $r$-band resolution, and the black dashed line divides the leading and trailing halves. Images for the rest of the jellyfish galaxy sample can be found in the Appendix.}
    \label{fig:panel_imgs}
\end{figure*}

Our parent sample of jellyfish galaxies is taken from \citet{roberts2021_LOFARclust}.  From a large sample of star-forming galaxies in low redshift clusters ($z < 0.05$), \citet{roberts2021_LOFARclust} identify 95 jellyfish galaxies on the basis of one-sided radio continuum tails observed by LOFAR at $144\,\mathrm{MHz}$ as part of the LOFAR Two Metre Sky Survey (LoTSS, \citealt{shimwell2022}). We match these 95 jellyfish galaxies against the final data release from the MaNGA survey (SDSS DR17, \citealt{abdurrouf2022}) that includes public IFS for $\sim\!10\,000$ galaxies in the main galaxy sample. Of the 95 jellyfish galaxies from \citet{roberts2021_LOFARclust}, 30 have MaNGA IFS.  Sixteen of these 30 galaxies are part of the Coma cluster, 10/30 are part of the A2197/A2199 system, 2/30 are part of the NGC6338 group, 1/30 is part of the NGC5098 group, and 1/30 is part of A2593. In Fig~\ref{fig:panel_imgs} (left-hand column) we show optical plus LOFAR overlay images for the galaxies that make up our final jellyfish galaxy sample.
\par
For these 30 galaxies we isolate star-forming spaxels according to emission line diagnostic diagrams \citep{baldwin1981}. We first mask all spaxels with $\mathrm{S/N} < 3$ in $\mathrm{H\alpha}$, $\mathrm{H\beta}$, [\textsc{Oiii}], [\textsc{Nii}], or [\textsc{Sii}], we then mask all spaxels with $\mathrm{EW(H\alpha}) < 3\,\mathrm{\AA}$ and finally mask any spaxels that are not classified as ``star-forming'' according to either the [\textsc{Oiii}]/$\mathrm{H\beta}$ versus [\textsc{Nii}]/$\mathrm{H\alpha}$ or [\textsc{Oiii}]/$\mathrm{H\beta}$ versus [\textsc{Sii}]/$\mathrm{H\alpha}$ dividing lines from \citet{kewley2001,kewley2006}. We note that this does not exclude spaxels that would be classified as ``composite'' between the \citet{kewley2001} and \citet{kauffmann2003} dividing lines on the [\textsc{Oiii}]/$\mathrm{H\beta}$ versus [\textsc{Nii}]/$\mathrm{H\alpha}$ diagram. We did re-run our analysis with a stricter selection for star-forming spaxels where we only include spaxels that have $\mathrm{EW(H\alpha}) > 6\,\mathrm{\AA}$ and added a requirement that spaxels are classified as ``star-forming'' according to the \citet{kauffmann2003} dividing line on the [\textsc{Oiii}]/$\mathrm{H\beta}$ versus [\textsc{Nii}]/$\mathrm{H\alpha}$ diagram. This did not qualitatively change any of the conclusions from this work, therefore we opt for the slightly looser criteria that includes composite spaxels to avoid over-masking weakly star-forming regions. This star-forming spaxel selection results in one galaxy being completely removed from our sample (MaNGA plate-ifu: 11943-6104) as all of the detected spaxels for this galaxy are classified as Seyfert AGN; however, the remaining 29 galaxies show emission dominated by star formation with a small minority of spaxels masked due to this selection. With non-star-forming spaxels masked we produce resolved SFR surface density maps, $\Sigma_\mathrm{SFR}$, for each jellyfish galaxy based on the MaNGA $\mathrm{H\alpha}$ emission line maps. We correct the $\mathrm{H\alpha}$ line fluxes for dust extinction using the Balmer decrement determined for each spaxel (see the appendix of \citealt{vogt2013} for a detailed description). The SFR for each spaxel is calculated from the dust-corrected $\mathrm{H\alpha}$ luminosity using the calibration from \citet{kennicutt1998_rev}. We then convert the SFR to a surface density in $\mathrm{M_\odot\,yr^{-1}\,kpc^{-2}}$ using the MaNGA spaxel dimensions of $0.5'' \times 0.5''$ and the distance to the host cluster. We obtain resolved stellar mass maps from the Pipe3D value added catalogs \citep{sanchez2016a,sanchez2016b,lacerda2022}. We use the stellar mass maps with dust corrections and apply the ``dezonification'' map to each stellar mass map in order to reduce the signature of the Voronoi binning (see \citealt{cidfernandes2014} for details). After multiplying by the dezonification map we follow \citet{sanchez2016b} and smooth each stellar mass map with a Gaussian kernel with a FWHM equal to the MaNGA $r$-band FWHM ($\sim\!2''-3''$, depending on the galaxy). Again, we convert the stellar mass maps to surface densities, $\Sigma_\star$, with units of $\mathrm{M_\odot\,kpc^{-2}}$. All spaxels with $\mathrm{S/N} < 3$ as well as all spaxels with $\Sigma_\star < 10^7\,\mathrm{M_\odot\,kpc^{-2}}$ in the stellar mass maps are masked.  Both SFR and stellar mass surface density maps are inclination corrected using the optical axis ratio ($b/a$) for each galaxy taken from the NASA-Sloan Atlas. Finally, we create resolved maps of specific star formation rate ($\mathrm{sSFR} = \mathrm{SFR} / M_\star$) by taking the ratio of the SFR and stellar mass surface density maps. In Fig.~\ref{fig:panel_imgs} (right-hand three columns) we show the SFR surface density, stellar mass surface density, and sSFR maps for each jellyfish galaxy.

\begin{figure}
    \centering
    \includegraphics[width=\columnwidth]{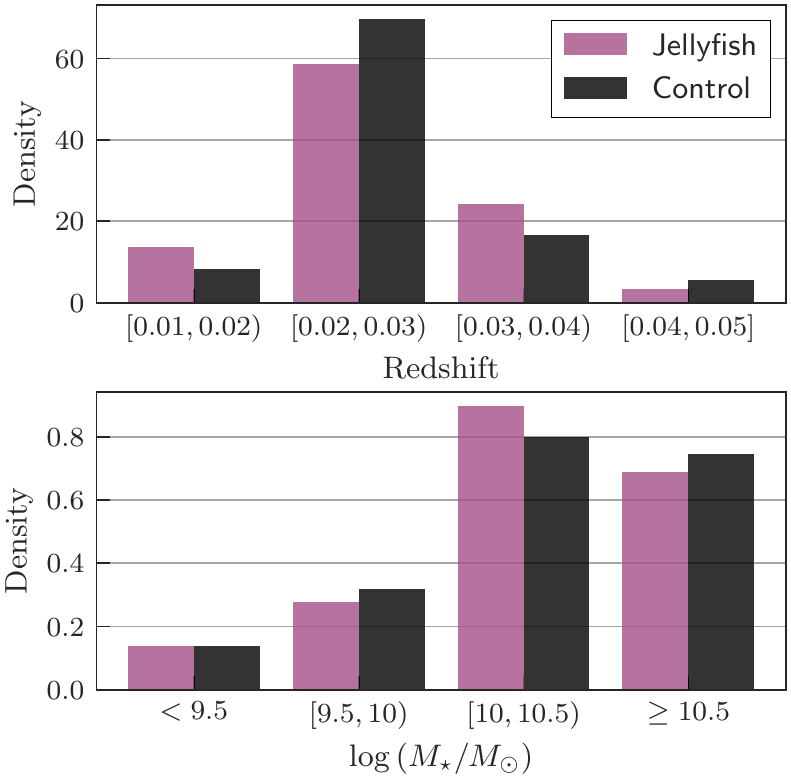}
    \caption{Redshift and integrated stellar mass distributions for jellyfish galaxies (magenta) and galaxies in the control sample (black).}
    \label{fig:sample_dist}
\end{figure}

\subsection{Control Galaxy Sample} \label{sec:control_sample}

We also compile a sample of matched control galaxies that are not members of massive groups or galaxy clusters. Comparing our results for jellyfish galaxies to the same properties measured for our control sample is critical for isolating the influence of the cluster environment, and in turn the process of RPS. We note that even galaxies in that are not in massive groups or clusters (e.g.\ galaxy pairs) can experience ``environmental-like'' effects on their star formation \citep[e.g.][]{moon2019,yang2022}. It is possible that cases such as this exist in our control sample, but the implicit assumption that we are making is that these types of perturbations will be significantly smaller than those experienced by satellite galaxies in massive clusters.  To generate our matched control sample we first consider all star-forming galaxies (i.e.\ $\mathrm{sSFR} > 10^{-11}\,\mathrm{yr^{-1}}$) in the DR17 MaNGA release \citep{abdurrouf2022}. We then restrict this to only those galaxies that are isolated (i.e.\ part of a single-member ``group'') or part of a group with halo mass, $M_H$, $<\!10^{13}\,\mathrm{M_\odot}$ in the \citet{lim2017} SDSS group/cluster catalogue. This ensures that our control sample does not contain galaxies in clusters or massive groups. We also restrict this sample to only contain galaxies that are detected by LOFAR at $144\,\mathrm{MHz}$ as part of the LoTSS DR2 source catalogue \citep{williams2019,shimwell2022}. This matches the selection criteria for the LOFAR jellyfish galaxy sample, which by definition are detected at 144 MHz.
\par
At this point we have an ``unmatched'' control sample of $\sim\!800$ star-forming MaNGA galaxies in low-density environments that have significant radio continuum emission at 144 MHz. We now match this sample to the stellar mass and redshift distribution of the jellyfish galaxy sample. By matching in stellar mass and redshift we ensure that the comparison between the jellyfish and control samples is unbiased in terms of differences in galaxy mass and physical resolution. For each jellyfish galaxy we find all galaxies in the unmatched control sample where: 
\begin{enumerate}
\item the control galaxy and jellyfish galaxy stellar masses are consistent within their respective $1\sigma$ error ranges

\item the difference between the control galaxy and jellyfish galaxy redshifts is given by $\Delta z < 0.005$
\end{enumerate}
\noindent
This gives a list of matches for each jellyfish galaxy, some jellyfish galaxies having $\sim\!5$ matched galaxies and some having tens of matched galaxies. To ensure that all jellyfish galaxies have equal weight in the final matched control sample, we randomly draw five matched control galaxies for each jellyfish galaxy giving a final control sample of 145 galaxies that are well matched to the jellyfish galaxy sample. The redshift and stellar mass distributions for both the jellyfish and control galaxy samples are shown in Fig.~\ref{fig:sample_dist}.

\section{Enhanced Star Formation on the Leading Half} \label{sec:enhanced_sfr}

\begin{figure*}
    \centering
    \includegraphics[width=\textwidth]{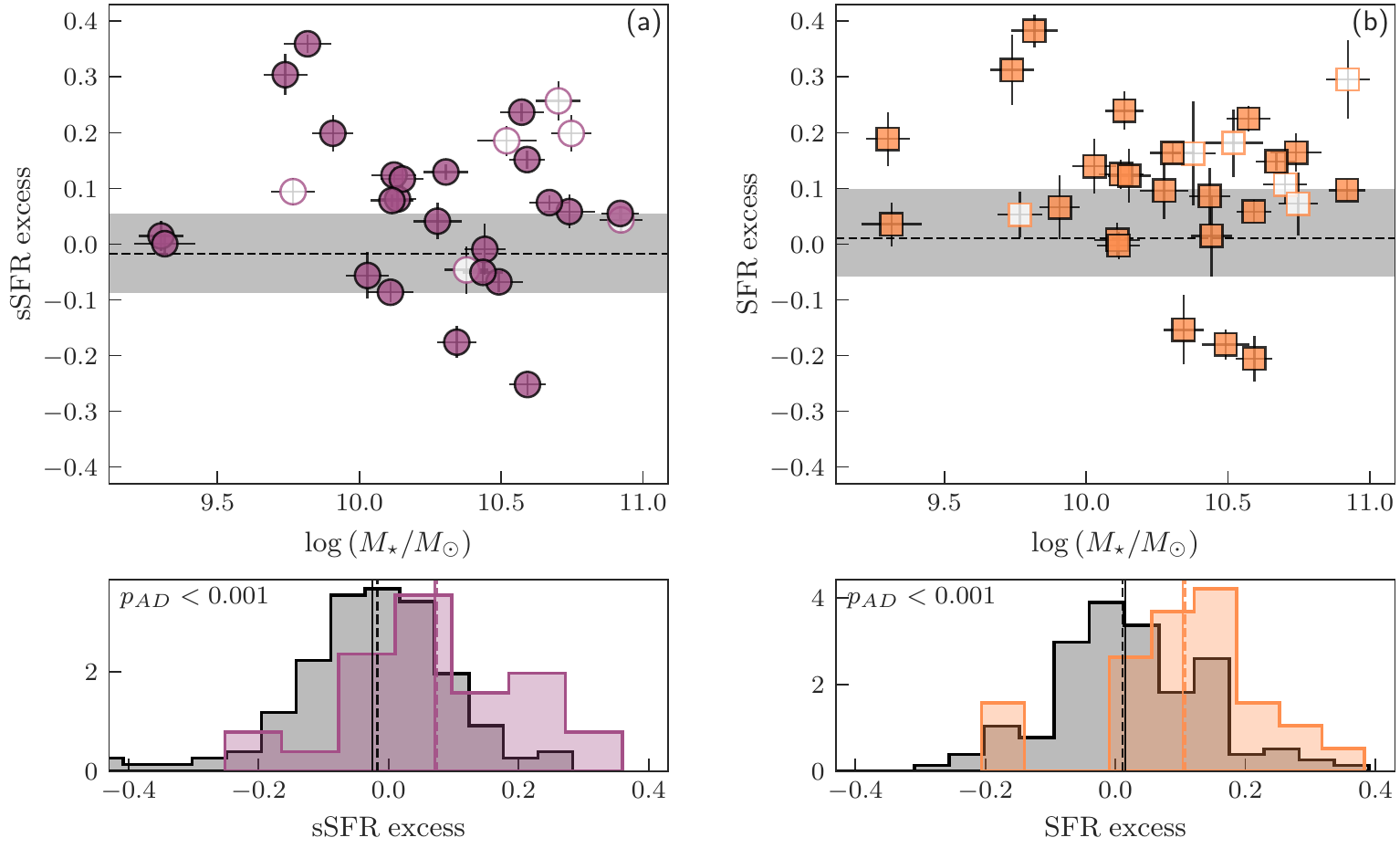}
    \caption{\emph{Left, top:} sSFR excess (see text for details) versus galaxy stellar mass. Data points correspond to the jellyfish galaxy sample, error bars on $\Delta \mathrm{sSFR}$ are calculated from bootstrap resampling and errorbars on stellar masses are taken from the Pipe3D catalog. The open markers correspond to the six jellyfish galaxies with marginally resolved $\Sigma_\mathrm{SFR}$ maps (along the axis of the observed tail), since the sSFR/SFR excess measurements for these galaxies are less reliable than the rest of the sample. The dashed line shows the median $\Delta \mathrm{sSFR}$ for the control sample and the shaded region encloses the interquartile range. \emph{Left, bottom:} Distributions of the sSFR excess for the jellyfish galaxy and control samples. Median (dashed) and mean (solid) are shown with the vertical lines. \emph{Right:} Same as left-hand panel but for the SFR excess (see text for details).}
    \label{fig:delta_sfr}
\end{figure*}

A primary question that we address in this work is whether or not star formation is enhanced on the leading half of jellyfish galaxies (ie.\ the galaxy side opposite to the direction of the RPS tail). Enhanced star formation on the leading half of group and cluster galaxies in the EAGLE simulations was reported by \citet{troncoso-iribarren2020}, which they argue is a result of gas compression in the ISM due to ram pressure. This work is an observational test of this prediction from hydrodynamical simulations.

\subsection{Star Formation Anisotropy} \label{sec:sfr_anisotropy}

\begin{deluxetable}{lccc}
\tabletypesize{\footnotesize}
\tablecaption{Jellyfish galaxy sSFR and SFR excess \label{tab:sfr_excess}}
\tablehead{
\colhead{Name} & \colhead{Plate-ifu} & \colhead{sSFR excess} & \colhead{SFR excess}
}
\startdata
LEDA43712 & 11006-3702 & $-0.09 \pm 0.02$ & $0.01 \pm 0.02$ \\
MRK0053 & 8934-3701 & $0.3 \pm 0.04$ & $0.31 \pm 0.04$ \\
IC3913 & 8932-12701 & $0.13 \pm 0.01$ & $0.16 \pm 0.01$ \\
IC3949 & 8950-12705 & $0.06 \pm 0.03$ & $0.16 \pm 0.03$ \\
IC0837 & 11009-12704 & $-0.07 \pm 0.02$ & $-0.18 \pm 0.02$ \\
GMP4351 & 9876-3702 & $-0.06 \pm 0.04$ & $0.14 \pm 0.04$ \\
MRK0056 & 11014-3704 & $0.2 \pm 0.03$ & $0.07 \pm 0.03$ \\
MRK0057 & 8932-3701 & $0.12 \pm 0.01$ & $0.13 \pm 0.01$ \\
$^*$GMP3509 & 8931-3703 & $0.09 \pm 0.02$ & $0.05 \pm 0.02$ \\
GMP3271 & 9876-3703 & $0.01 \pm 0.03$ & $0.19 \pm 0.03$ \\
GMP2601 & 9863-12701 & $0.0 \pm 0.02$ & $0.04 \pm 0.02$ \\
GMP2599 & 9862-9101 & $0.08 \pm 0.02$ & $0.24 \pm 0.02$ \\
GMP1616 & 8935-6104 & $-0.18 \pm 0.03$ & $-0.15 \pm 0.03$ \\
GMP1576 & 9876-6101 & $0.12 \pm 0.02$ & $0.12 \pm 0.02$ \\
GMP0455 & 11004-12701 & $0.36 \pm 0.02$ & $0.38 \pm 0.02$ \\
$^*$LEDA2017338 & 8442-1901 & $-0.05 \pm 0.04$ & $0.16 \pm 0.04$ \\
$^*$LEDA2168096 & 12673-6101 & $0.19 \pm 0.03$ & $0.18 \pm 0.03$ \\
$^*$LEDA2175783 & 11942-6103 & $0.26 \pm 0.04$ & $0.11 \pm 0.04$ \\
LEDA58067 & 8603-9102 & $0.07 \pm 0.01$ & $0.15 \pm 0.01$ \\
MRK0881 & 8604-9102 & $-0.25 \pm 0.02$ & $-0.21 \pm 0.02$ \\
LEDA2156782 & 8550-3701 & $-0.01 \pm 0.04$ & $0.01 \pm 0.04$ \\
LEDA58296 & 8312-12703 & $0.06 \pm 0.01$ & $0.1 \pm 0.01$ \\
$^*$LEDA2147644 & 9869-12702 & $0.04 \pm 0.02$ & $0.3 \pm 0.02$ \\
$^*$LEDA58307 & 9869-9102 & $0.2 \pm 0.03$ & $0.07 \pm 0.03$ \\
UGC10429 & 8550-6103 & $0.15 \pm 0.02$ & $0.06 \pm 0.02$ \\
LEDA2568088 & 8625-9102 & $0.08 \pm 0.01$ & $-0.0 \pm 0.01$ \\
LEDA2566358 & 8625-12702 & $0.24 \pm 0.02$ & $0.23 \pm 0.02$ \\
LEDA1479719 & 8622-6103 & $-0.05 \pm 0.02$ & $0.09 \pm 0.02$ \\
\enddata
\tablecomments{$^*$ $\Sigma_\mathrm{SFR}$ map is only marginally resolved along the axis of the radio tail}
\end{deluxetable}

In \citet{troncoso-iribarren2020} the ``leading'' and ``trailing'' halves of galaxies are defined relative to the three-dimensional velocity vector for each galaxy, such that the plane that passes through the galaxy center and is normal to the velocity vector, divides the leading and trailing halves. This is a natural choice as ram pressure is maximal along the direction of motion, but the three-dimensional velocity vector is not an observable quantity. We approximate this by using the direction of the observed radio continuum tail as a proxy for the direction of motion for each of the jellyfish galaxies in our sample. Tail directions are taken from \citet{roberts2021_LOFARclust} and range between $0^\circ$ and $360^\circ$, where $0^\circ = \mathrm{West}$ and $90^\circ = \mathrm{North}$. We assume that the radio continuum tail points opposite to the direction of motion and thus we observationally divide the leading and trailing half by the line passing through the optical galaxy center that is normal to the projected tail direction. For each galaxy in our control sample we also define a leading and trailing half by dividing the galaxy with a line through the optical galaxy center at a random orientation. This division is not physically motivated but it does quantify the level of intrinsic anisotropy expected between two halves of a ``normal'' galaxy not part of an overdense environment. This allows us to isolate the impact of environment for our jellyfish sample by determining whether anisotropies between the leading and trailing halves of jellyfish galaxies systematically differ from those found by randomly dividing galaxies in our control sample.
\par
With a leading half and trailing half defined for each galaxy in our jellyfish and control samples, we now define two quantities, the sSFR excess and the SFR excess, to measure the degree of star formation anisotropy between these two galaxy halves. We calculate these quantities as:
\begin{equation}
    \mathrm{sSFR\;excess} = \log_{10}\left(\,\mathrm{\overline{sSFR}_{LH}}\,\middle/\, \mathrm{\overline{sSFR}_{TH}}\, \right),
\end{equation}
\noindent and
\begin{equation}
    \mathrm{SFR\;excess} = \log_{10}\left(\,\mathrm{\overline{SFR}_{LH}}\,\middle/\, \mathrm{\overline{SFR}_{TH}}\, \right),
\end{equation}
\noindent
where $\mathrm{\overline{sSFR}_{LH}}$ and $\mathrm{\overline{sSFR}_{TH}}$ are the average sSFRs measured over the leading half and trailing half of the galaxy and $\mathrm{\overline{SFR}_{LH}}$ and $\mathrm{\overline{SFR}_{TH}}$ are defined analogously for SFR. Positive values of the sSFR and SFR excess indicate enhanced star formation on the leading half relative to the trailing half and vice versa for $\mathrm{sSFR/SFR\;excess} < 0$. We list the measured sSFR excess and SFR excess values for each jellyfish galaxy in Table~\ref{tab:sfr_excess}. There are six jellyfish galaxies in our sample (MaNGA plate-ifus: 8931-3703, 8442-1901, 12673-6101, 11942-103, 9869-12702, 9869-9102) where only $\sim\!2$ resolution elements (MaNGA $r$-band FWHM) span across the $\Sigma_\mathrm{SFR}$ map along the axis of the observed radio tail. These are primarily edge-on galaxies with stripped tails that are roughly aligned with the galaxy minor axis. For these galaxies, the distinction between leading half and trailing half is only marginally resolved, and thus the measurement of sSFR/SFR excess is likely less reliable than for the rest of the sample. We have tested removing these galaxies from the sample and re-running our analysis and we confirm that the qualitative results from this work do not change based on the inclusion or exclusion of these six galaxies. The median sSFR/SFR excess for the jellyfish sample is 0.07/0.11 when including these poorly resolved cases and 0.06/0.10 when excluding them.  We opt to include these galaxies in our analysis but highlight them with open markers in Figs~\ref{fig:delta_sfr}, \ref{fig:delta_ang}, \ref{fig:max_ssfr} and with asterisks in Table~\ref{tab:sfr_excess}.
\par
In Fig.~\ref{fig:delta_sfr}a we show the sSFR excess and in Fig.~\ref{fig:delta_sfr}b we show the SFR excess, both plotted against integrated galaxy stellar mass for galaxies in our jellyfish sample. We also show the median sSFR and SFR excess for our control sample (dashed line) as well as the interquartile range (shaded regions). Jellyfish galaxies are skewed to positive values for both the sSFR and SFR excess, this is clear both in the upper panels of Fig.~\ref{fig:delta_sfr} as well as lower panels where we show the mean/median (solid/dashed vertical lines) values and the full distributions of sSFR and SFR excess for both the jellyfish and the control samples. For the control sample the sSFR and SFR excess is centered on zero, which is expected given that we are defining the leading half and trailing half along a random axis for these galaxies. There is statistical evidence that the distributions of the sSFR excess and SFR excess for jellyfish galaxies and galaxies in the control sample are distinct. According to the two-sample Anderson-Darling test \citep{scholz1987} this difference is significant at $>$99.9 per cent significance for both sSFR and SFR. While the distributions of sSFR and SFR excess for jellyfish galaxies are skewed to positive values and consistent with enhanced star formation on the leading half, not all jellyfish galaxies show this signature. A ram pressure induced burst of star formation is likely a transient phenomenon and thus we do not expect to observe all jellyfish galaxies in this state. It is also possible that for some galaxies there is a clear star formation excess signal in 3D that is being obscured by projection effects. Furthermore the degree to which the ISM is perturbed by ram pressure will depend on a given galaxy's orbital history and position within its host cluster. We explore whether jellyfish galaxies with enhanced star formation on the leading half occupy a distinct part of projected phase space relative to jellyfish galaxies without strong SFR anisotropies, but do not find any significant difference. If we divide our sample into two sub-samples split at the median value of the SFR excess (the results are similar for sSFR excess) then the median and interquartile range of the clustercentric radius (i.e.\ $R/R_{180}$ as given in \citealt{roberts2021_LOFARclust}) distribution is 0.41 [0.22,0.53] for galaxies with large excesses and 0.37 [0.28,0.48] for galaxies with small excesses. Doing the same for velocity offset (i.e.\ $c\Delta z / \sigma$ as given in \citealt{roberts2021_LOFARclust}) gives 0.69 [0.45,1.95] for galaxies with large excesses and 0.75 [0.40,1.55] for galaxies with small excesses. According to the two-sample Anderson-Darling test there is no statistical evidence for a difference between the phase space distributions for galaxies with large and small star formation excesses, but we note that with our modest sample size of 29 jellyfish galaxies it is difficult to subdivide the sample in such a way while maintaining statistical power.
\par
\citet{troncoso-iribarren2020} compute an analogous quantity to our SFR excess for the simulated group/cluster galaxies in their work (equation 8 in \citealt{troncoso-iribarren2020}). In these simulations the SFR for each gas particle is set according to the observed KS-relation, with a metallicity-dependent density threshold to prevent the unrealistic case of hot, low density gas forming stars. We note that these SFR excesses are measured in a relative sense within individual galaxies and this adds some degree of robustness in terms of uncertainties surrounding the specifics of star formation prescriptions between observations and simulations. In their simulated data, \citet{troncoso-iribarren2020} find enhanced star formation on the leading half with a typical SFR excess of $\sim\!0.05$. They show that their simulated group/cluster galaxies have enhanced gas pressures on their leading sides (both relative to the trailing sides and relative to EAGLE main sequence galaxies). This increased pressure feeds into their SFR prescription (equation 1 in \citealt{troncoso-iribarren2020}) and thus this SFR excess is attributed to this gas compression on the leading side.  We find a typical SFR excess  of $\sim\!0.1$ for the observed jellyfish galaxies in this work, slightly larger than is found by \citeauthor{troncoso-iribarren2020}. This contrast may be explained by difference in the sample definitions. In particular, \citet{troncoso-iribarren2020} include all star-forming group/cluster galaxies from the simulation box in their analysis, whereas we are restricted to only star-forming galaxies with an observed radio continuum tail. Our jellyfish galaxy sample is likely skewed towards stronger RPS in order to produce an observable tail, which in turn may drive a stronger enhancement of star formation on the leading half. Furthermore, our jellyfish sample is drawn from massive clusters ($M_\mathrm{halo} > 10^{14}\,\mathrm{M_\odot}$) whereas the galaxy sample in \citet{troncoso-iribarren2020} has a substantial number of galaxies hosted by group-mass systems ($M_\mathrm{halo} < 10^{14}\,\mathrm{M_\odot}$). \citet{roberts2021_LOFARgrp} and \citet{roberts2022_UNIONS} have shown that global SFR enhancements in galaxies undergoing RPS are stronger in massive clusters and more marginal in group-mass systems. All said, the qualitative results between this work and \citet{troncoso-iribarren2020} are in good agreement, as both works find evidence for enhanced star formation on the leading half of galaxies likely due to RPS.

\subsection{Maximizing the Star Formation Anisotropy} \label{sec:max_anisotropy}

\begin{figure*}
    \centering
    \includegraphics[width=\textwidth]{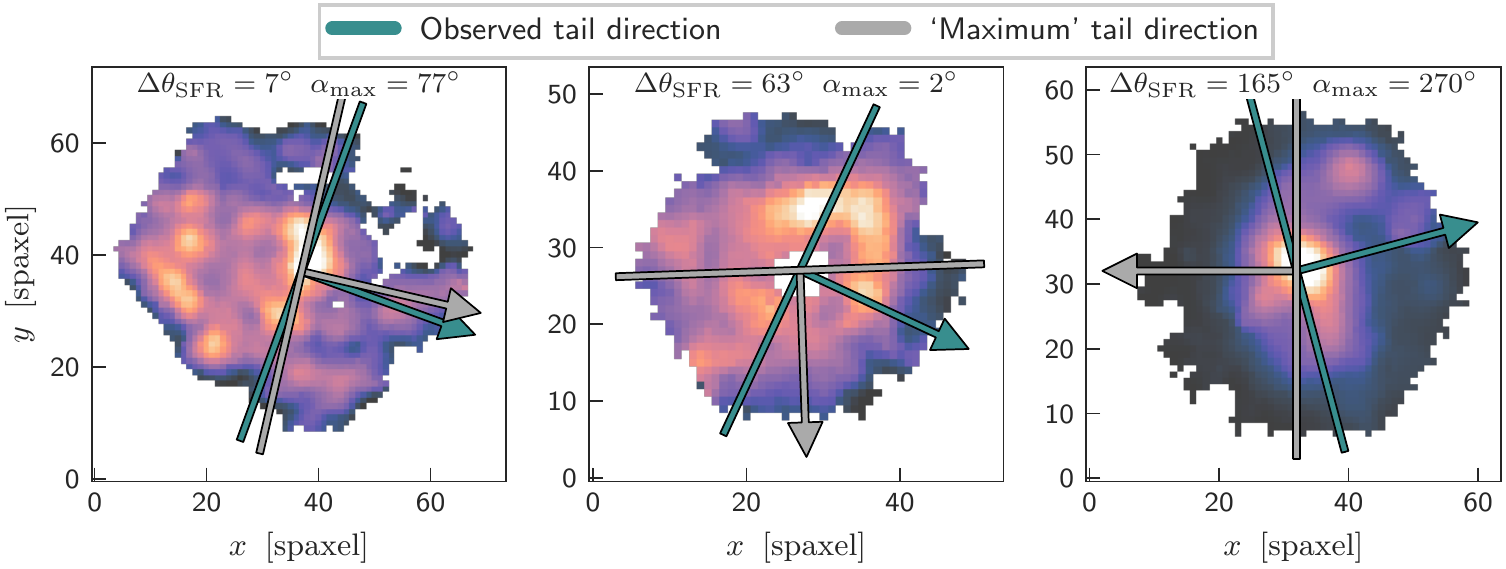}
    \caption{Example schematics showing the observed tail direction (teal arrow) and the ``tail direction'' that is inferred by maximizing the SFR excess (grey arrow), overlaid on the SFR surface density maps. The solid lines show the corresponding divisions between the leading and trailing sides. We show three examples, one with $\Delta \theta_\mathrm{SFR} \sim 0$ (IC3913), one with an intermediate value of $\Delta \theta_\mathrm{SFR}$ (UGC10429), and one with $\Delta \theta_\mathrm{SFR} \sim 180$ (MRK0881).}
    \label{fig:tail_defs}
\end{figure*}

\begin{figure}
    \centering
    \includegraphics[width=\columnwidth]{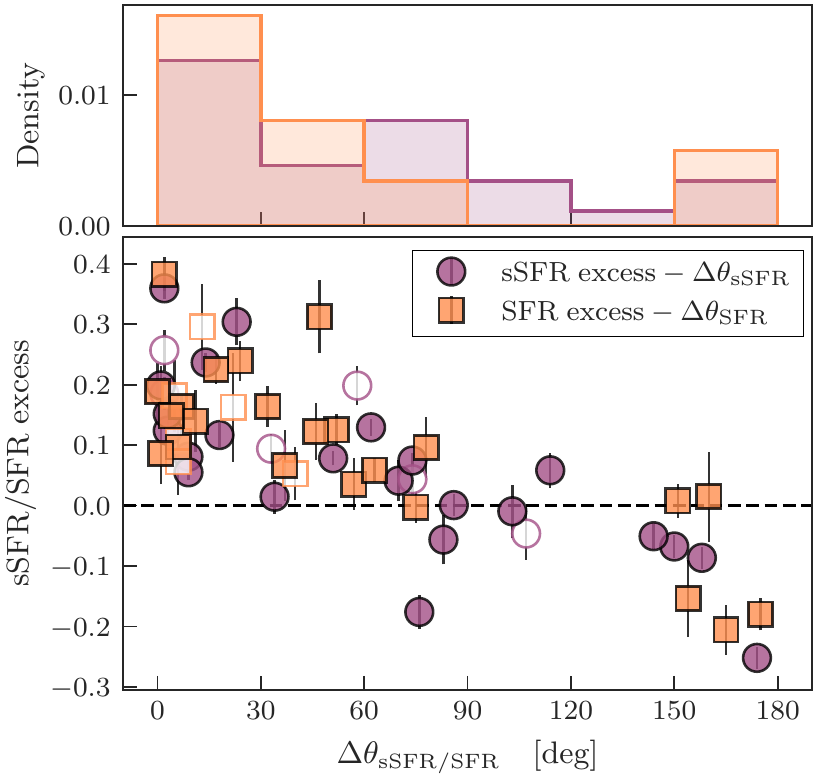}
    \caption{sSFR/SFR excess versus the offset between the axis which maximizes the star formation anisotropy and the leading-trailing half axis inferred from the observed tail direction. Star formation anisotropies measured with the sSFR excess are shown with the magenta circles and star formation anisotropies measured with the SFR excess are shown with orange squares. Errorbars on the sSFR/SFR excess are calculated from bootstrap resampling. The open markers correspond to the six jellyfish galaxies with marginally resolved $\Sigma_\mathrm{SFR}$ maps (along the axis of the observed tail). In the upper panel we also show a histogram of the distributions for $\Delta \theta_\mathrm{sSFR}$ and $\Delta \theta_\mathrm{SFR}$.}
    \label{fig:delta_ang}
\end{figure}

In the previous section we show that our sample of jellyfish galaxies show evidence for enhanced star formation on their leading halves, opposite to the direction of the RPS tail. We define the leading half and trailing half of our galaxies according to the direction of observed radio continuum tails, though an alternative approach to quantify star formation anisotropy in galaxies is with a dividing line that maximizes the sSFR or SFR excess. In the case of ram pressure induced star formation from gas compression, these two definitions should be similar. In other words, if ram pressure is driving enhanced star formation, then the dividing line that maximizes the sSFR or SFR excess should be roughly normal to the observed tail direction (modulo projection effects). Maximizing this star formation anisotropy is also suggested by \citet{troncoso-iribarren2020} as a viable method for determining the leading half and trailing half of group/cluster galaxies observationally, which may be particularly useful in cases where a clear RPS tail is not observed.
\par
In this section we test how closely our definition of the leading half and trailing half (based on the observed tail directions) corresponds to the dividing line that maximizes the sSFR and SFR excess for each galaxy. For each jellyfish galaxy we first find two maximal dividing lines: one to maximize the sSFR excess and one to maximize the SFR excess.  All dividing lines must pass through the optical galaxy center and are otherwise defined by an orientation, $\alpha$, where $\alpha \in [0^\circ,360^\circ)$. An orientation of $0^\circ$ corresponds to a horizontal line along the W-E axis with the northern side of the galaxy as the leading half and an orientation of $90^\circ$ corresponds to a vertical line along the N-S axis with the eastern side of the galaxy as the leading half. From the orientation that maximizes the sSFR and SFR excess, $\alpha_\mathrm{max,\,sSFR}$ and $\alpha_\mathrm{max,\,SFR}$, we then infer a ``tail direction'' that is normal to this angle and can be compared to the observed tail direction. In Fig.~\ref{fig:tail_defs} we show these different tail direction definitions graphically and in Fig.~\ref{fig:delta_ang} we plot the sSFR/SFR excess versus the difference between the observed tail direction and the direction that is normal to the line maximizing the sSFR excess ($\Delta \theta_\mathrm{sSFR}$) and the SFR excess ($\Delta \theta_\mathrm{SFR}$). A value of $0^\circ$ corresponds to the case where the tail direction inferred from maximizing the sSFR/SFR excess is identical to the observed tail direction, for $90^\circ$ the two are perpendicular, and for $180^\circ$ the two point in opposite directions. In Fig.~\ref{fig:delta_ang} there is a clear anti-correlation between both the sSFR excess and $\Delta \theta_\mathrm{sSFR}$ as well as the SFR excess and $\Delta \theta_\mathrm{SFR}$. This shows that jellyfish galaxies with large sSFR/SFR excess also have observed tail directions that roughly maximize the difference in star formation between the leading half and trailing half -- in agreement with expectations from RPS.  Jellyfish galaxies with $\mathrm{sSFR/SFR\;excess} \sim 0$ have values of $\Delta \theta_\mathrm{sSFR}$ and $\Delta \theta_\mathrm{SFR}$ that span the full range between $0^\circ$ and $180^\circ$ and thus do no appear to be tied to the observed tail direction. Finally, there is a hint that the few galaxies with large negative sSFR/SFR excesses tend to have $\Delta \theta_\mathrm{sSFR/SFR}$ near $180^\circ$. Even in these cases the orientation that maximizes the sSFR/SFR excess seems to be connected to the tail direction. This may be an indicator of star formation occurring in the stripped gas being transported along the tail direction, and we also briefly discuss this possibility in the Section~\ref{sec:peak_ssfr}. If strong star formation anistropies are observed in cluster galaxies, these results suggest that it may be possible to infer a RPS tail direction by maximizing this anisotropy (as also suggested by \citealt{troncoso-iribarren2020}). That said, such an exercise should always be taken with significant caution as there are physical mechanisms in cluster environments beyond RPS that can produce asymmetric star formation.

\subsection{Locating the Peak of (Specific) Star Formation} \label{sec:peak_ssfr}

\begin{figure}
    \centering
    \includegraphics[width=\columnwidth]{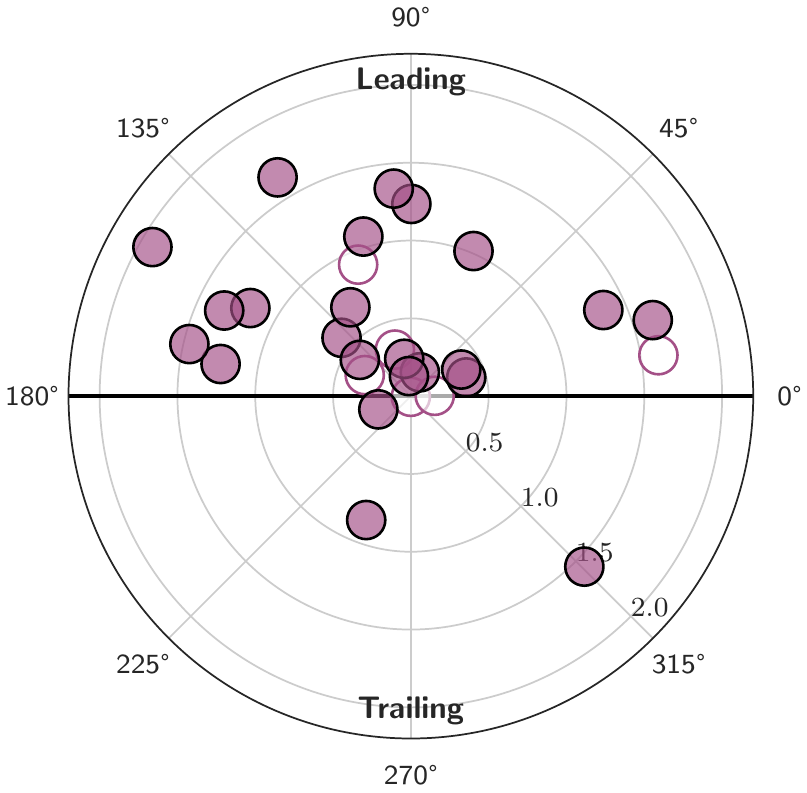}
    \caption{The location of the specific star formation rate peak for each jellyfish galaxy in polar coordinates. The azimuthal axis shows the orientation between the sSFR peak and the observed tail (see text for details) and the radial axis shows the radial offset of the sSFR peak from the galaxy center in units of the galaxy effective radius. The open markers correspond to the six jellyfish galaxies with marginally resolved $\Sigma_\mathrm{SFR}$ maps (along the axis of the observed tail).}
    \label{fig:max_ssfr}
\end{figure}

In Sections~\ref{sec:sfr_anisotropy} and \ref{sec:max_anisotropy} we have shown that jellyfish galaxies have enhanced sSFRs on their leading half and that the dividing line between the leading half and trailing half inferred from observed tail directions is close to the dividing line that maximizes the star formation anisotropy for the majority of jellyfish galaxies in our sample. As a final test to confirm enhanced star formation on the leading half of these galaxies we explore the position of the sSFR peak for each galaxy with respect to the observed radio continuum tail. We use the peak of sSFR instead of SFR since most galaxies will have a SFR peak near the galaxy center due to the high concentration of mass, whereas by using sSFR we are probing a relative increase in SFR modulated by the local stellar mass surface density.
\par
We identify the position of the pixel with the highest sSFR, $p_\mathrm{max} = (x_\mathrm{max}, y_\mathrm{max})$, and compare the location of $p_\mathrm{max}$ to the observed direction of the stripped tail. When determining $p_\mathrm{max}$ we filter the sSFR map with a $3 \times 3$ uniform kernel such that each pixel is averaged with its eight nearest neighbors. This is to ensure that our measurement is robust against random pixel-to-pixel varations. We quantify the orientation between the two with an angle such that $p_\mathrm{max}$ on the leading half of the galaxy will have an orientation between $0^\circ$ and $180^\circ$ (where $90^\circ$ is directly opposite to the tail direction) and $p_\mathrm{max}$ on the trailing half will have an orientation between $180^\circ$ and $360^\circ$ (with $270^\circ$ being directly in line with the tail). For brevity we will refer to this orientation as $\Delta \varphi$. We also record the radial offset of $p_\mathrm{max}$ with respect to the galaxy center in units of the galaxy effective radius, which we denote as $r_\mathrm{max}$. In Fig.~\ref{fig:max_ssfr} we plot $\Delta \varphi$ (azimuthal axis) and $r_\mathrm{max}$ (radial axis) on a polar plot for each jellyfish galaxy. The distribution of points is asymmetric with the majority of jellyfish galaxies having $0 < \Delta \varphi < 180$, demonstrating the that the regions of peak star formation (per unit stellar mass) are systematically found on the leading half. The two galaxies in Fig.~\ref{fig:max_ssfr} with sSFR peaks that are on the trailing half and significantly offset from the galaxy center are GMP2599 (MaNGA plate-ifu: 9862-9101) and MRK0881 (MaNGA plate-ifu: 8604-9102).  These sSFR peaks may be indicative of ongoing star formation along the stripped tail for these galaxies as has been seen previously for some jellyfish galaxies \citep[e.g.][]{gavazzi2001,yagi2010,poggianti2017,boselli2018,hess2022,lee2022_gmos_tails}. The MaNGA field-of-view (FOV) is too small to directly probe these stripped tails; however, for both galaxies the $\mathrm{H\alpha}$ emission fills the MaNGA FOV toward the direction of the observed radio tail suggesting that there may be extra-planar star formation along the tail that continues beyond the MaNGA coverage.

\section{A Molecular Gas Case Study: IC3949} \label{sec:ic3949}

\begin{figure}
    \centering
    \includegraphics[width=\columnwidth]{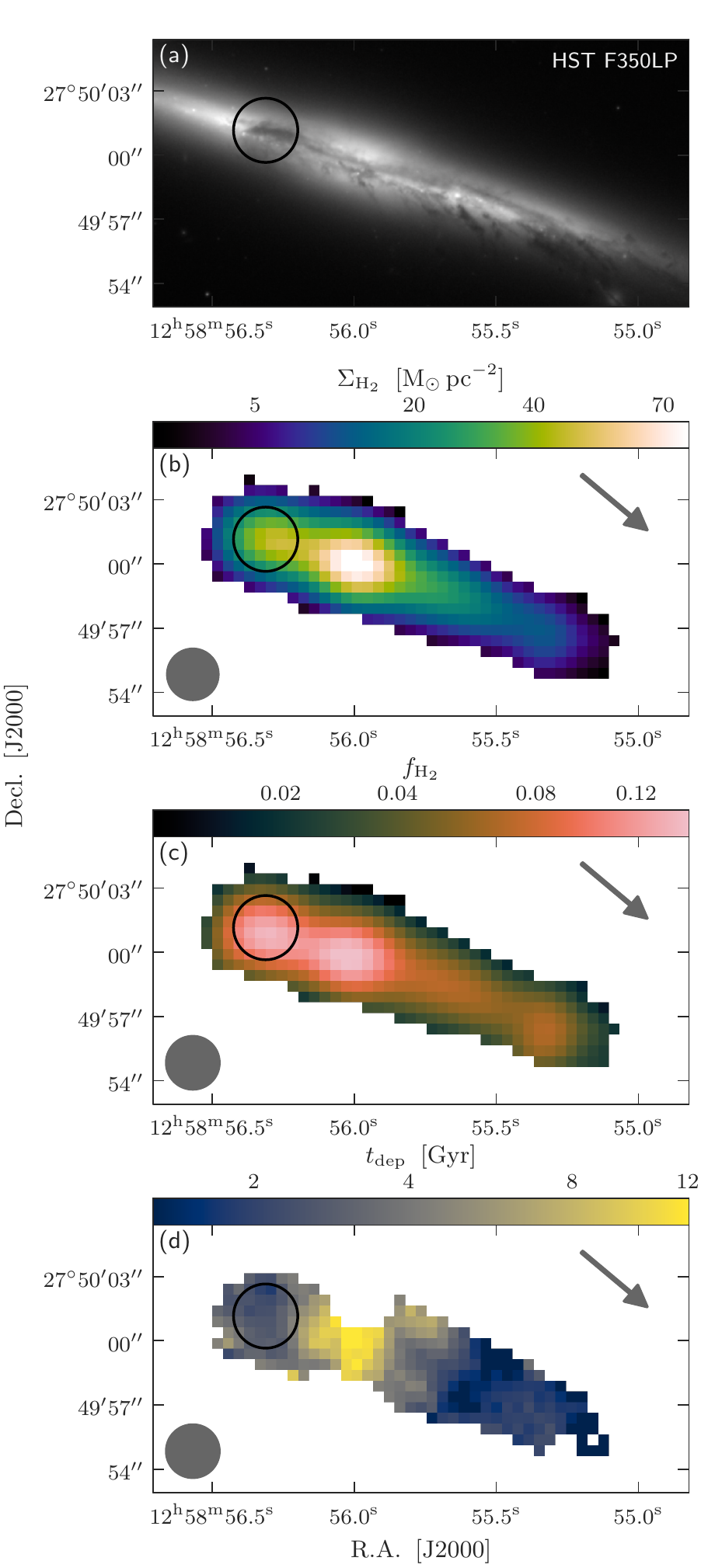}
    \caption{Maps of molecular gas and star formation properties in IC3949. From top to bottom: \emph{Hubble Space Telescope} near-UV (F350LP), $\mathrm{H_2}$ surface density, $\mathrm{H_2}$ gas fraction ($\Sigma_\mathrm{H_2} / \Sigma_\star$), and $\mathrm{H_2}$ depletion time ($\Sigma_\mathrm{H_2} / \Sigma_\mathrm{SFR}$). The arrow in each panel shows the orientation of the observed RPS tail and the filled circle shows the FWHM resolution for the ALMA/MaNGA products. As a visual guide we also highlight the leading side region of enhanced star formation in each panel with the open circle. The \textit{HST} data used in this figure can be found in MAST: \dataset[10.17909/7ftv-0w17]{http://dx.doi.org/10.17909/7ftv-0w17}}
    \label{fig:ic3949_co}
\end{figure}

IC3949 (MaNGA plate-ifu: 8950-12705) is a massive ($\log [M_\star/M_\odot] \simeq 10.7$), edge on disk galaxy located at the center of the Coma Cluster ($R/R_{180}=0.1$, \citealt{roberts2021_LOFARclust}) that is undergoing RPS. IC3949 shows a stripped tail to the southwest that was first identified by \citet{yagi2010} through narrowband $\mathrm{H\alpha}$ observations. This tail has since been confirmed in the radio continuum at both 144 MHz (\citealt{roberts2021_LOFARclust}) and 350 MHz \citep{lal2022}. IC3949 is seemingly at an advanced quenching stage as it shows a highly truncated $\mathrm{H\alpha}$ disk (relative to the optical extent of the galaxy) and a low integrated SFR for its stellar mass ($\mathrm{SFR \approx 0.7\,M_\odot\,yr^{-1}}$, \citealt{salim2016,salim2018}) that places it in the so-called ``green valley''. In Fig.~\ref{fig:ic3949_co}a we show a near-UV image (F350LP) of IC3949 taken by the \textit{Hubble Space Telescope}. This image shows UV continuum emission from the star-forming disk along with a complex dust morphology visible through extinction. Filaments of dust are visible extending to the south/southwest in the direction of the observed radio continuum tail, likely being stripped by ram pressure.  The most prominent region of dust extinction is co-spatial with the region of enhanced star formation on the leading side of the disk (see Fig.~\ref{fig:panel_imgs} and the open circle in Fig.~\ref{fig:ic3949_co}a).
\par
IC3949 is the one jellyfish galaxy in our sample with both resolved optical spectroscopy from MaNGA and public, resolved, molecular gas observations taken by the Atacama Large Millimeter/sub-millimeter Array (ALMA). For IC3949, this allows us to not only explore the star formation activity relative to the RPS tail but also probe the molecular gas density and depletion time as a function of position within this galaxy. IC3949 was observed in CO J=1-0 in 2018 (P.I. Lin, 2017.1.01093.S) and in HCN/HCO$^+$ in 2020 (P.I. Lin, 2019.1.01178.S) as part of the larger ALMA MaNGA Quenching and Star-Formation Survey (ALMaQUEST, \citealt{lin2020}).  Below we use these ALMA data to explore the molecular gas morphology in IC3949, in particular in the context of the region on enhanced star formation observed on the leading side of the galaxy.

\subsection{Bulk Molecular Gas Distribution} \label{sec:co_dist}

In this section we analyze the bulk molecular gas distribution in IC3949, as traced by CO J=1-0 emission. We use CO products that were reduced and imaged by the ALMaQUEST team. Here we provide a description of the key components of this data processing, for a more detailed outline of the ALMaQUEST reduction procedure please see \citet{lin2020}. IC3949 was observed in the C43-2 configuration of the ALMA-12m array with an on-source integration time of $\sim\!22$min. The data were calibrated using CASA 5.4 (Common Astronomy Software Applications, \citealt{mcmullin2007}) and the standard ALMA pipeline. The continuum was subtracted from the visibilities and CLEAN was used to clean the continuum subtracted data down to $1\sigma$. Brigg's weighting with a robust parameter of 0.5 was used for the imaging. In order to match the pixel scale and the typical resolution of MaNGA data products, a user-specified pixel size of $0.5''$ and a restoring beam size of $2.5''$ FWHM were set. The resulting spectral line cube for CO J=1-0 has a velocity channel width of $\sim\!11\,\mathrm{km\,s^{-1}}$ and an rms of $0.6\,\mathrm{mJy\,beam^{-1}}$. A CO moment-zero map for IC3949 was produced with \texttt{immoments} in CASA including only velocity channels where CO emission is present. We estimate the rms noise from source-free regions in the CO moment-zero map and any pixels with $\mathrm{S/N < 3}$ are masked. The CO moment-zero map is converted to an $\mathrm{H_2}$ surface density by adopting a constant conversion factor of $\alpha_\mathrm{CO} = 4.35\,\mathrm{M_\odot\,(K\,km\,s^{-1}\,pc^{-2})^{-1}}$ \citep[e.g.][]{bolatto2013}, and this map is shown in Fig.~\ref{fig:ic3949_co}b. In Fig.~\ref{fig:ic3949_co} we also show maps of $\mathrm{H_2}$ gas fraction and depletion time that are obtained by dividing the $\mathrm{H_2}$ surface density map by the stellar mass and SFR surface density maps, respectively (see Fig~\ref{fig:panel_imgs} for the $\Sigma_\star$ and $\Sigma_\mathrm{SFR}$ maps for IC3949).  In each panel we show an arrow denoting the projected orientation of the observed radio continuum tail and an open circle highlighting the region of enhanced star formation on the leading half.

\begin{figure}
    \centering
    \includegraphics[width=\columnwidth]{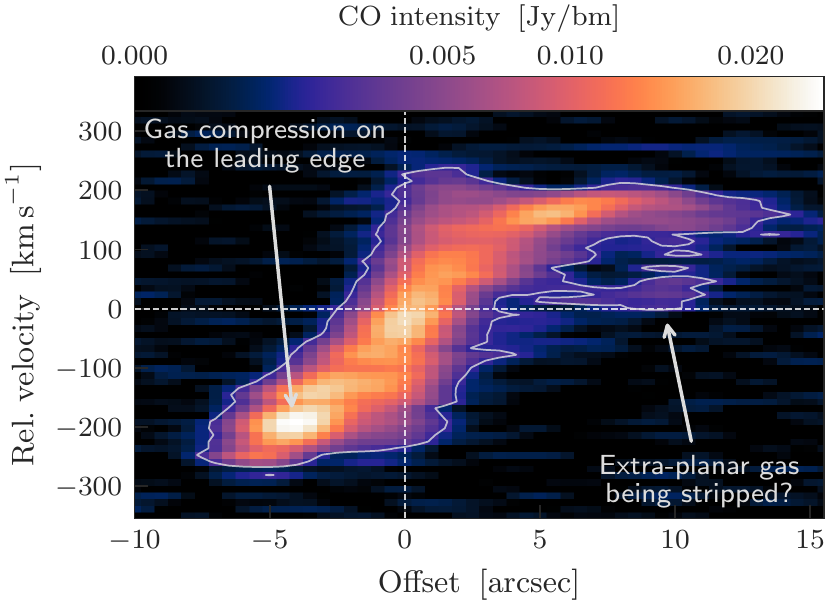}
    \caption{Position-velocity diagram (PVD) for IC3949 measured along the direction of the major axis (which roughly corresponds to the direction of the stripped tail). The width of the PVD box encompasses all of the CO emission shown in Fig.~\ref{fig:ic3949_co}. The grey contour corresponds to $3\times \mathrm{rms}$ of the CO cube.}
    \label{fig:ic3949_pvd}
\end{figure}

Relative to the galaxy center, the molecular gas disk of IC3949 is clearly asymmetric, being truncated on the leading half of the galaxy (opposite to the tail direction) and more extended along the trailing half. This can likely be attributed to RPS which is transporting gas from the leading side of IC3949 ``downstream'' along the direction of the stripped tail.  This ALMA observation does not detect any molecular gas in the stripped tail itself, as has been seen previously for some jellyfish galaxies \citep{jachym2014,jachym2017,jachym2019,moretti2020,moretti2020b}. It is worth noting that the observational set-up was designed for observing molecular gas in the disk therefore the primary beam response is substantially lower in the tail region beyond the optical extent of the galaxy. Forthcoming CO J=2-1 observations of IC3949 from the ALMA-JELLY survey (P.I.\ J{\'a}chym, 2021.1.01616.L) will place stronger constraints on the presence, or lack thereof, of molecular gas in the RPS tail of IC3949. There is molecular gas asymmetry in the disk of IC3949, not just in terms of extent, but also in terms of $\mathrm{H_2}$ surface density. The $\mathrm{H_2}$ surface density surface density is larger on the leading side of the disk (at fixed galactocentric radius), coincident with the enhanced star formation seen on the leading half. This enhanced CO emission is subtle in the moment-zero map (Fig.~\ref{fig:ic3949_co}a) but more clearly visible in both the $\mathrm{H_2}$ gas fraction map (Fig.~\ref{fig:ic3949_co}b) and the position-velocity diagram (PVD) shown in Fig.~\ref{fig:ic3949_pvd}. This enhanced $\mathrm{H_2}$ surface density is consistent with a framework where ram pressure compresses gas on the leading side of galaxies, in turn catalyzing increased star formation \citep[e.g.][]{gavazzi2001,cramer2020,boselli2021_ic3476,cramer2021,roberts2022_perseus}. The PVD also shows a faint collection of extra-planar gas located at an offset of $\sim\!10''$ along the trailing side of the galaxy. This gas is detached from the CO rotation curve and is potentially indicative of a plume of molecular gas being directly stripped from the disk of IC3949.
\par
The depletion time is not constant across the disk of IC3949 (see Fig.~\ref{fig:ic3949_co}c) but instead is longest at the galaxy center and decreases radially outwards, ranging between $1\,\mathrm{Gyr}$ and $11\,\mathrm{Gyr}$. Previous studies find a mixture of enhanced or suppressed depletion times in the centers of late-type galaxies with the origin of these variations still being debated \citep[e.g.][]{utomo2017,colombo2018,chown2019}. \citet{utomo2017} find, on average, shorter central gas depletion times in the EDGE-CALIFA sample of galaxies and argue that gas compression from a higher central stellar potential may shorten the depletion time. In contrast to this, \citet{chown2019} find that it is primarily barred galaxies in the EDGE-CALIFA sample that have shorter central depletion times, while unbarred galaxies appear to have longer depletion times relative to their disks. In the centre of IC3949, star formation appears suppressed relative to the molecular gas content, despite appearing to have a higher gas surface density. It is unclear what drives this suppression, but one explanation could be increased turbulence \citep[e.g.][]{salas2021} that is potentially connected to the ongoing RPS. The spaxels within the galaxy center of IC3949 are classified as ``composite'' according to the resolved BPT diagram from MaNGA, thus it is possible that low-level AGN activity is contributing to the $\mathrm{H\alpha}$ flux in this region. This, however, will not resolve the difference between the central depletion time relative to the rest of the galaxy as this would actually imply that we are overestimating the central SFR and thus underestimating the depletion time. Finally, it is important to note that there is some evidence that $\alpha_\mathrm{CO}$ is lower in galaxy centers \citep[e.g.][]{sandstrom2013}, which in turn would decrease the observed central depletion time. Over the region of enhanced star formation on the leading half of IC3949 (circular aperture shown in Fig.~\ref{fig:ic3949_co}) the average depletion time is $2.7\,\mathrm{Gyr}$, identical to the median value over the whole galaxy. Thus the enhanced star formation does not result from a high star formation efficiency but instead an increase in molecular gas that is in line with the resolved Kennicutt-Schmidt (rKS, \citealt{schmidt1959,kennicutt1998}) relation for IC3949.

\subsection{Dense Molecular Gas Distribution} \label{sec:dense_gas}

\begin{figure}
    \centering
    \includegraphics[width=\columnwidth]{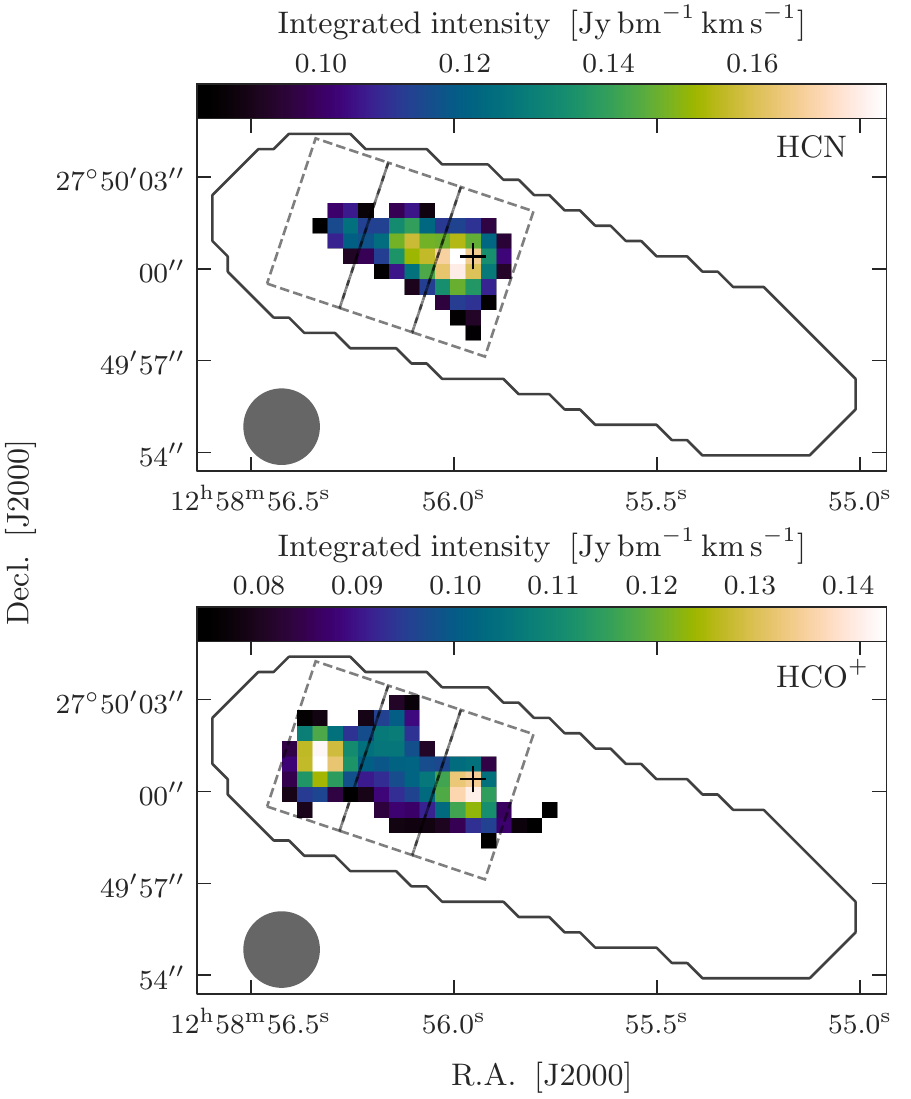}
    \caption{Integrated intensity maps of HCN and HCO$^+$ for IC3949. The solid contour shows the extent of the CO emission in IC3949, the cross marks the optical galaxy center, and the filled circle shows the FWHM beam size.}
    \label{fig:dense_gas_mom0}
\end{figure}

\begin{figure}
    \centering
    \includegraphics[width=\columnwidth]{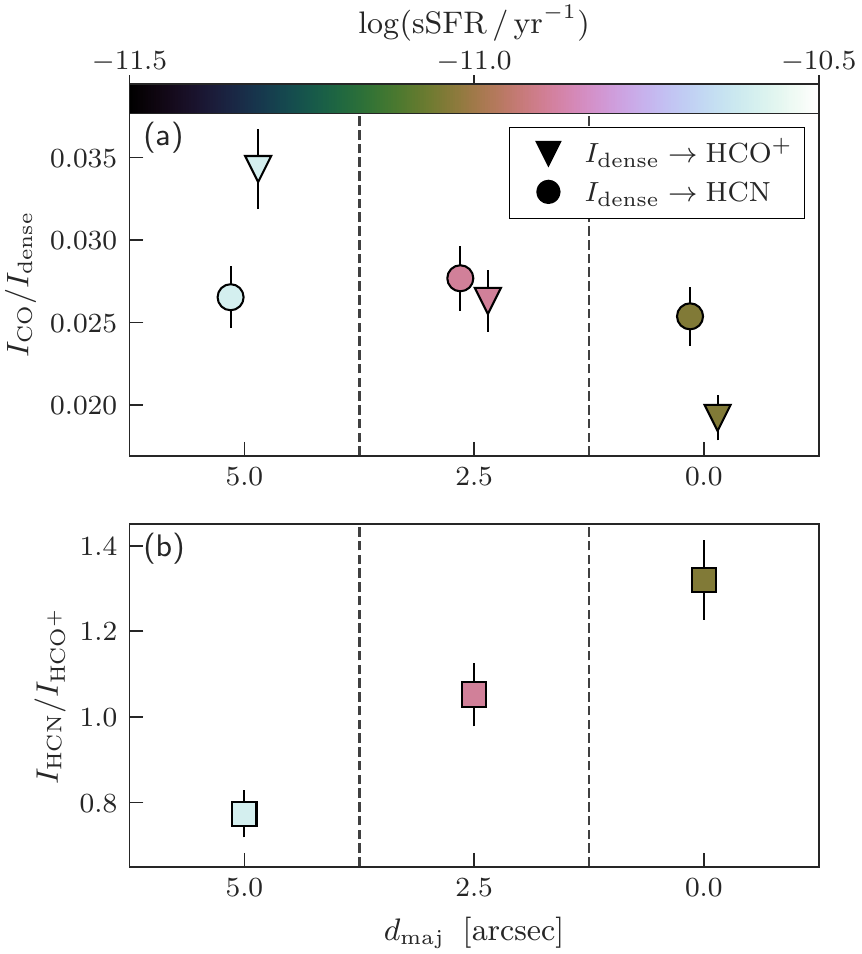}
    \caption{\emph{Top:} Dense gas fraction ($I_\mathrm{dense} / I_\mathrm{CO}$) derived from HCN (circles) and HCO$^+$ (triangles) within the three apertures shown in Fig~\ref{fig:dense_gas_mom0}. The x-axis, $d_\mathrm{maj}$, corresponds to the distance of the aperture center along the galaxy major axis (in the direction of the leading edge). \emph{Bottom:} HCN--HCO$^+$ line ratio within the three apertures shown in Fig~\ref{fig:dense_gas_mom0}. In both panels the data points are colored by the average sSFR in each aperture and the error bars are a combination of random error derived from bootstrap resampling and a 5\% calibration uncertainty assumed for each line flux (ALMA Band 3, \href{https://almascience.eso.org/proposing/technical-handbook}{ALMA Technical Handbook}).}
    \label{fig:dense_gas_fractions}
\end{figure}

For a subset of ALMaQUEST galaxies, including IC3949, follow up observations of the J=1-0 transitions for the dense gas molecular tracers HCN and HCO$^+$ were obtained in ALMA Cycle 7. HCN and HCO$^+$ have critical densities of $\sim\!10^5\,\mathrm{cm^{-3}}$ and $\sim\!5\times10^4\,\mathrm{cm^{-3}}$ \citep{shirley2015}, respectively, compared to $\sim\!10^{3}\,\mathrm{cm^{-3}}$ for CO, and thus trace a denser component of the ISM.  Here we describe and present the imaging of the HCN and HCO$^+$ lines for IC3949. A full analysis of the five ALMaQUEST galaxies with dense gas observations will be presented in a forthcoming paper (Lin et al., in prep.).
\par
Dense gas observations at $88.631\,\mathrm{GHz}$ (PI Lin, 2019.1.01178.S) were carried out with ALMA in Cycle 7 using the Band 3 receiver and C43-2 configuration. These observations cover the full FOV of the CO J=1-0 observations described in the previous section.  The spectral setup includes one line targeting HCN J=1-0 and three low-resolution spectral windows for the continuum. The line spectral window has a bandwidth of $\sim\!930\,\mathrm{MHz}$ ($3200\,\mathrm{km\,s^{-1}}$), with a native channel width of $\sim\!2\,\mathrm{km\,s^{-1}}$, and is sufficiently wide to also include the HCO$^{+}$ J=1-0 line ($89.189\,\mathrm{GHz}$). The data were processed by the standard pipeline in CASA 5.6. Continuum is subtracted from the data in the visibility domain. The task \texttt{tclean} was employed for deconvolution with a Briggs robust parameter of 0.5. We adopted a user-specified image center, pixel size ($0.5''$), and restoring beamsize ($2.5''$) to match the image grid and the spatial resolution of the MaNGA images. The restoring beamsize is similar to that of the native beamsize reported by the \texttt{tclean} ($2.7'' \times 2.4''$).  
The  HCN and HCO$^{+}$ lines were imagined separately and to increase the signal-to-noise ratio the spectral channels are binned to $\sim\!50\,\mathrm{km\,s^{-1}}$. The rms noise of the spectral line data cubes are $0.13\,\mathrm{mJy\,beam^{-1}}$ and $0.14\,\mathrm{mJy\,beam^{-1}}$ for HCN and HCO$^+$, respectively. The integrated intensity maps of HCN and HCO$^{+}$ were constructed using the task \texttt{immoments} in CASA by integrating emission from a velocity range set by hand to match the observed line profile without any clipping in signal. 
\par
In Fig.~\ref{fig:dense_gas_mom0} we show integrated intensity (moment-zero) maps of HCN and HCO$^+$ for IC3949. We also show the outline of the detected CO emission from Fig.~\ref{fig:ic3949_co}a with the solid contour and the galaxy center with the cross. Emission in the moment zero maps (for both HCN and HCO$^+$) is restricted to the galaxy center and the leading half of IC3949, with no significant emission detected on the trailing half. The fact that emission from these dense gas tracers is only detected along the leading side is additional evidence for gas compression from ram pressure as was argued for in Section~\ref{sec:co_dist}. There are also morphological differences between HCN and HCO$^+$ in Fig.~\ref{fig:dense_gas_mom0}. HCN emission peaks at the galaxy center and decreases radially towards the leading edge of the galaxy, whereas HCO$^+$ emission is bimodal with a peak at the galaxy center and a second peak that is spatially coincident with the region of enhanced star formation on the leading half. We also measure the intensity ratio between dense gas tracers and CO, $I_\mathrm{dense}/I_\mathrm{CO}$
where $I_\mathrm{dense} = I_\mathrm{HCN}$ or $I_\mathrm{dense} = I_\mathrm{HCO^+}$, and the HCN-to-HCO$^+$ ratio ($I_\mathrm{HCN}/I_\mathrm{HCO^+}$) in three rectangular apertures that span the detected dense gas emission in the moment zero maps (see Fig.~\ref{fig:dense_gas_mom0}). These apertures are oriented along the major axis of IC3949, are spaced in intervals of $2.5''$, i.e.\ one beam's width, and cover the area between the galaxy center ($d_\mathrm{maj} = 0''$ in Fig.~\ref{fig:dense_gas_fractions}) and the region of enhanced star formation on the leading edge of IC3949 ($d_\mathrm{maj} = 5''$ in Fig.~\ref{fig:dense_gas_fractions}). We color the data points in Fig.~\ref{fig:dense_gas_fractions} by the average sSFR in each aperture in order to further emphasize the increase in star formation towards the leading edge. As shown in Fig.~\ref{fig:dense_gas_fractions}a, the dense gas fraction traced by HCO$^+$ increases monotonically toward the region of enhanced star formation on the leading side, whereas the dense gas fraction traced by HCN remains constant across all three apertures. These differences between HCN and HCO$^+$ are further illustrated in Fig.~\ref{fig:dense_gas_fractions}b where we plot the HCN-to-HCO$^+$ ratio measured in each of the three apertures. This ratio is largest at the galaxy center and decreases monotonically toward the region of enhanced star formation on the leading side, where HCO$^+$ becomes brighter than HCN and the ratio falls below unity.  Previous works have found that low HCN-HCO$^+$ ratios are typically associated with starburst regions in galaxies \citep[e.g.][]{kohno2001,meijerink2007,privon2015,bemis2019,butterworth2022,zhou2022}. This is likely due to an increase in HCO$^+$ abundance driven by the enhanced UV radiation fields and cosmic ray ionization expected in regions of strong star formation \citep[e.g.][]{krips2008,meijerink2011,nayana2020}. Our resolved observations of the HCN-to-HCO$^+$ ratio in IC3949 supports such a picture as the region of enhanced star formation on the leading side is characterized by a particularly low HCN-to-HCO$^+$ ratio. Though we also note that HCO$^+$ has a lower critical density than HCN \citep[e.g.][]{shirley2015}, thus the HCN-HCO$^+$ ratio may be decreasing away from the galaxy center, in part, due to a decreasing average density of molecular gas.

\section{Discussion and Conclusions} \label{sec:discussion_conclusions}

In this work we use MaNGA IFS observations to present evidence for enhanced star formation on the leading half for a sample of 29 jellyfish galaxies. This highlights a key phase experienced by galaxies undergoing RPS where ram pressure is capable of enhancing star formation prior to quenching. Thus the SFR does not necessarily decrease monotonically through the environmental quenching process. The fact that this SFR enhancement is connected to the orientation of the stripped tail (see Fig.~\ref{fig:delta_sfr}, \ref{fig:delta_ang}) lends further credence to this enhancement being connected to gas compression from ram pressure. This is in line with the compression that is apparent in the LOFAR radio emission where steep surface brightness gradients (i.e.\ closely spaced contour levels) are visible on the leading edges for many of these jellyfish galaxies (see Fig.~\ref{fig:panel_imgs}). The results from this work are in agreement with previous observational studies that have demonstrated enhanced star formation on the leading edge of select galaxies undergoing RPS \citep{gavazzi2001,boselli2021_ic3476,roberts2022_perseus,hess2022} as well as the results from the EAGLE hydrodynamical simulation by \citet{troncoso-iribarren2020}. More broadly this work is also consistent with the fact that, on a population level, galaxies undergoing RPS have integrated SFRs that are sytematically enhanced \citep[e.g.][]{dressler1983,ebeling2014,vulcani2018_sf,roberts2020,wang2020,roberts2021_LOFARclust,durret2021,lee2022_enhanced_sfr}. Based on the results of this work it is likely that these enhanced integrated SFRs are being primarily driven by star formation activity on the leading side. This work presents the largest sample of galaxies for which ram pressure enhanced star formation has been resolved and studied, though this may be extended in the near future. Upon completion of the Canada-France Imaging Survey \citep{ibata2017}, deep, high-resolution $u$-band imaging will be available as a resolved SFR tracer for the entire sample of $\sim\!150$ LOFAR jellyfish galaxies from \citet{roberts2021_LOFARgrp,roberts2021_LOFARclust}, expanding the sample size relative to this work by a factor of $\sim\!5$.
\par
In Section~\ref{sec:ic3949} we present a case study on the distribution of molecular gas for one jellyfish galaxy, IC3949, in our sample. For IC3949 there is evidence for an increase in molecular gas density over the observed region of enhanced star formation on the leading half.  The relationship between $\mathrm{H_2}$ surface density and SFR in this region follows the expectation from the rKS relation for IC3949, suggesting that the enhanced star formation is a result of an increase in molecular gas (likely due to compression from ram pressure) and not an increase in SFE. We note that \citet{tomicic2018} present evidence for short depletion times (high SFE) on the leading side of the nearby group galaxy NGC2276 that is undergoing RPS (and also likely a tidal interaction with NGC2300).  NGC2276 is a starburst galaxy ($\mathrm{SFR} \approx 10-20\,\mathrm{M_\odot\,yr^{-1}}$, \citealt{kennicutt1983,wolter2015,tomicic2018}) whereas IC3949 is in the green valley at an advanced quenching stage ($\mathrm{SFR \approx 0.7\,M_\odot\,yr^{-1}}$, \citealt{salim2016,salim2018}), thus this difference may reflect the very different stages of star formation for these two galaxies. \citet{villanueva2022} study a sample of 38 Virgo galaxies from the VERTICO survey \citep{brown2021} and find that the SFE within the effective radius decreases and the molecular-to-atomic gas ratio increases with increasing environmental perturbation (as traced by HI morphologies and deficiencies). They attribute this to environmental effects that remove atomic gas from galaxies, drive molecular gas to the central regions of galaxies, and/or promote the conversion from atomic to molecular gas. The median SFE of $3.7 \times 10^{-10}\,\mathrm{yr^{-1}}$ ($t_\mathrm{dep} = 2.7\,\mathrm{Gyr}$) that we measure for IC3949 is consistent with the range of values that \citet{villanueva2022} find for Virgo galaxies with perturbed HI morphologies. There is further observational evidence for ram pressure promoting the formation of molecular gas \citep[e.g.][]{moretti2020,moretti2020b}, and in particular driving enhanced molecular gas densities along the leading edge of galaxies undergoing RPS \citep[e.g.][]{lee2017,cramer2020,cramer2021}. The enhanced $\mathrm{H_2}$ density that we show in Fig.~\ref{fig:ic3949_pvd} appears similar to the region of gas compression that \citet{cramer2020} highlight in the PVD for the Virgo cluster galaxy NGC4402.  Additionally, the dense molecular gas tracers HCN and HCO$^+$ are only detected in IC3949 over a region from the galaxy center along the leading side -- not on the trailing half of the galaxy. This reinforces the notion that there are increased ISM densities on the leading half.  To our best knowledge, Section~\ref{sec:dense_gas} is the first ever analysis of the distribution of dense molecular gas tracers in a jellyfish galaxy undergoing RPS. Moving forward, dense gas observations will hopefully be obtained for more galaxies undergoing RPS in order to elucidate whether the signatures observed for IC3949 are typical or not.

\subsection*{Summary}

In this work we present clear evidence for an enhancement of star formation on the leading half of galaxies undergoing ram pressure stripping. This star formation enhancement is consistent with predictions from hydrodynamical simulations and is likely driven by gas compression via ram pressure. Below we highlight the main scientific conclusions from this work.
\begin{enumerate}
    \item Jellyfish galaxies show enhanced SFRs and sSFRs on the leading half of the galaxy, opposite to the direction of the stripped tail. (Fig.~\ref{fig:delta_sfr}).
    
    \item The dividing line that maximizes the star formation anisotropy for each jellyfish galaxy is systematically aligned with the dividing line between the leading and trailing halves of the galaxy as inferred from the observed tail direction (Fig.~\ref{fig:delta_ang}).
    
    \item For the jellyfish galaxies in our sample, the location of the maximum sSFR is preferentially found on the leading half of the galaxy (Fig.~\ref{fig:max_ssfr}).
    
    \item The jellyfish galaxy IC3949 has an area of enhanced molecular gas surface density on the leading side of the galaxy that is spatially coincident with the observed star formation enhancement. This region has a typical $\mathrm{H_2}$ depletion time (as traced by CO) compared to the rest of the galaxy, suggesting that the enhancement star formation is driven by an increase in molecular gas and not an abnormally high star formation efficiency (Figs~\ref{fig:ic3949_co} and \ref{fig:ic3949_pvd}).
    
    \item In IC3949, emission from dense molecular gas tracers HCN and HCO$^+$ is only detected between the galaxy center and the leading edge, reinforcing the increase in ISM density on the leading side (Fig.~\ref{fig:dense_gas_mom0}).
\end{enumerate}
\noindent
Recent and forthcoming ALMA large programs such as VERTICO \citep{brown2021} and ALMA-JELLY (P.I.\ J{\'a}chym, 2021.1.01616.L) will significantly increase the number of galaxies undergoing RPS with resolved CO observations and thus shed further light on the relationship between molecular gas and star formation in such systems. We find evidence for ram pressure compression of the ISM in IC3949, but it is still not clear whether this is a generic feature of all galaxies undergoing RPS. These large CO surveys will take a significant step towards addressing this open question and will also provide a parent sample from which compelling targets can be selected for follow-up observations of dense gas tracers and/or CO at higher spatial resolution. For a more complete understanding of the impact of RPS on star formation and molecular gas, probing the ISM at cloud-scales in jellyfish galaxies should be a priority moving forward.

\section*{Acknowledgments}
\noindent We thank anonymous referee for their constructive comments on the paper. We also thank the Leiden/ESA Astrophysics Program for Summer Students for hosting DT while working on this project. IDR and RJvW acknowledge support from the ERC Starting Grant Cluster Web 804208. LL acknowledges support from the Academia Sinica under the Career Development Award CDA107-M03 and the Ministry of Science and Technology of Taiwan under the grant MOST 111-2112-M-001-044. HAP acknowledges support from the National Science and Technology Council of Taiwan under grant 110-2112-M-032-020-MY3. AI acknowledges the INAF founding program `Ricerca Fondamentale 2022' (PI A. Ignesti).
\\[0.5em]
Funding for the Sloan Digital Sky Survey IV has been provided by the Alfred P. Sloan Foundation, the U.S. Department of Energy Office of Science, and the Participating 
Institutions. SDSS-IV acknowledges support and resources from the Center for High Performance Computing  at the University of Utah. The SDSS website is \url{www.sdss.org}.
SDSS-IV is managed by the Astrophysical Research Consortium for the Participating Institutions of the SDSS Collaboration including the Brazilian Participation Group, the Carnegie Institution for Science, Carnegie Mellon University, Center for Astrophysics | Harvard \& Smithsonian, the Chilean Participation Group, the French Participation Group, Instituto de Astrof\'isica de Canarias, The Johns Hopkins University, Kavli Institute for the Physics and Mathematics of the Universe (IPMU) / University of Tokyo, the Korean Participation Group, Lawrence Berkeley National Laboratory, Leibniz Institut f\"ur Astrophysik Potsdam (AIP),  Max-Planck-Institut f\"ur Astronomie (MPIA Heidelberg), 
Max-Planck-Institut f\"ur Astrophysik (MPA Garching), Max-Planck-Institut f\"ur 
Extraterrestrische Physik (MPE), National Astronomical Observatories of China, New Mexico State University, New York University, University of Notre Dame, Observat\'ario Nacional / MCTI, The Ohio State University, Pennsylvania State University, Shanghai Astronomical Observatory, United Kingdom Participation Group, Universidad Nacional Aut\'onoma de M\'exico, University of Arizona, University of Colorado Boulder, University of Oxford, University of Portsmouth, University of Utah, University of Virginia, University of Washington, University of Wisconsin, Vanderbilt University, and Yale University.
\\[0.5em]
This paper makes use of the following ALMA data: \\[0.25em]
ADS/JAO.ALMA \href{https://almascience.nrao.edu/aq/?result_view=projects&projectCode=\%222017.1.01093.S\%22}{\#2017.1.01093.S} \\
ADS/JAO.ALMA \href{https://almascience.nrao.edu/aq/?result_view=projects&projectCode=\%222019.1.01178.S\%22}{\#2019.1.01178.S}. \\[0.25em]
ALMA is a partnership of ESO (representing its member states), NSF (USA) and NINS (Japan), together with NRC (Canada), MOST and ASIAA (Taiwan), and KASI (Republic of Korea), in cooperation with the Republic of Chile. The Joint ALMA Observatory is operated by ESO, AUI/NRAO and NAOJ. In addition, publications from NA authors must include the standard NRAO acknowledgement: The National Radio Astronomy Observatory is a facility of the National Science Foundation operated under cooperative agreement by Associated Universities, Inc.
\\[0.5em]
\emph{Software:} \textsc{astropy} \citep{astropy2013}, \textsc{casa} \citep{mcmullin2007}, \textsc{cmasher} \citep{vandervelden2020}, \textsc{ds9} \citep{ds9_2003}, \textsc{marvin} \citep{cherinka2019}, \textsc{matplotlib} \citep{hunter2007}, \textsc{numpy} \citep{harris2020}, \textsc{photutils} \citep{bradley2022}, \textsc{rsmf} (\url{https://rsmf.readthedocs.io/en/latest/source/howto.html}), \textsc{scipy} \citep{virtanen2020}, \textsc{spectral-cube} (\url{https://spectral-cube.readthedocs.io/en/latest/})


\bibliography{main}
\bibliographystyle{aasjournal}

\appendix

\section{Figure Set Images}
\label{sec:img_appendix}

\begin{figure*}
    \centering
    \includegraphics[width=\textwidth]{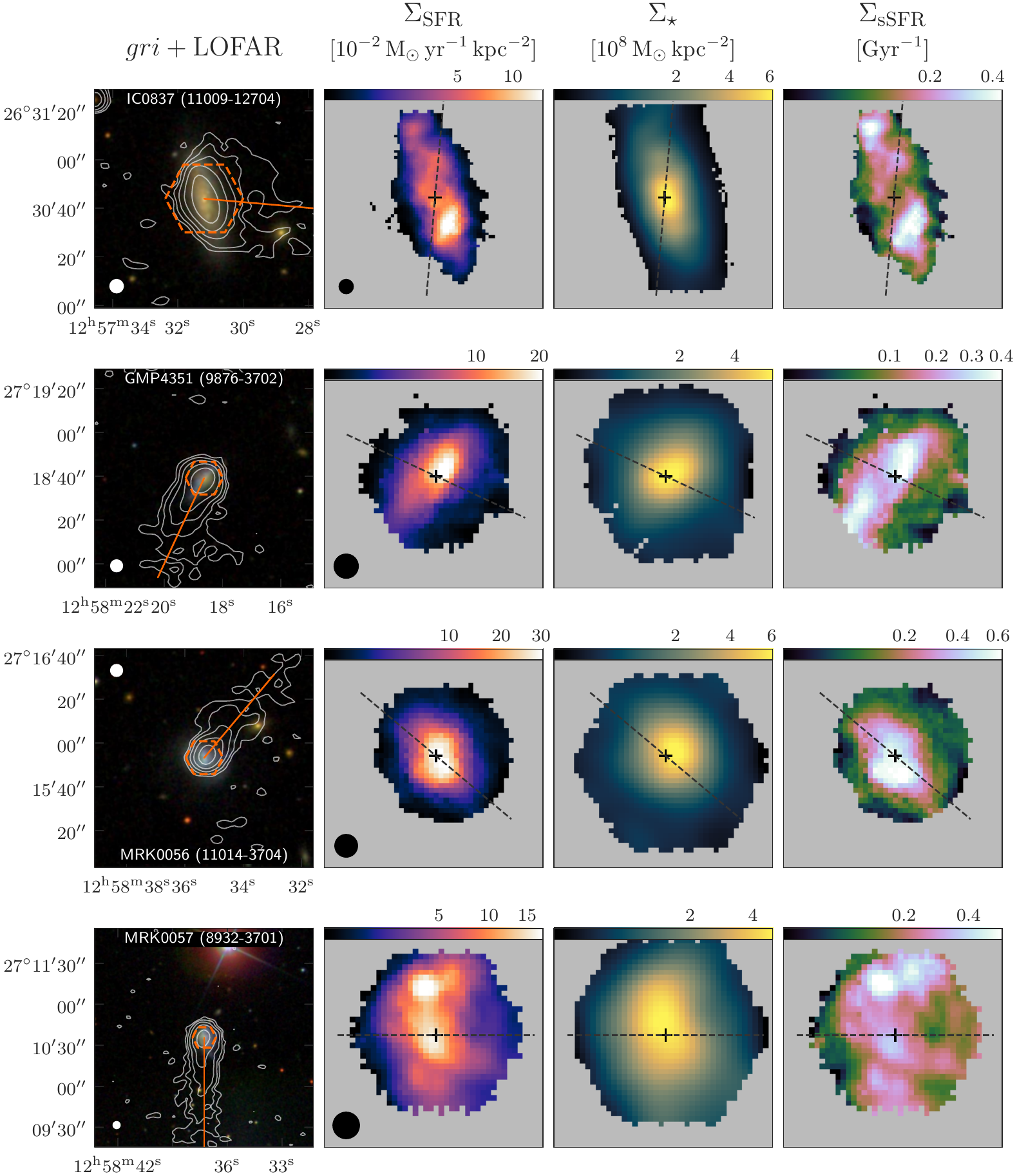}
    \caption{See Fig.~\ref{fig:panel_imgs}}
    \label{fig:panel_img1}
\end{figure*}

\begin{figure*}
    \centering
    \includegraphics[width=\textwidth]{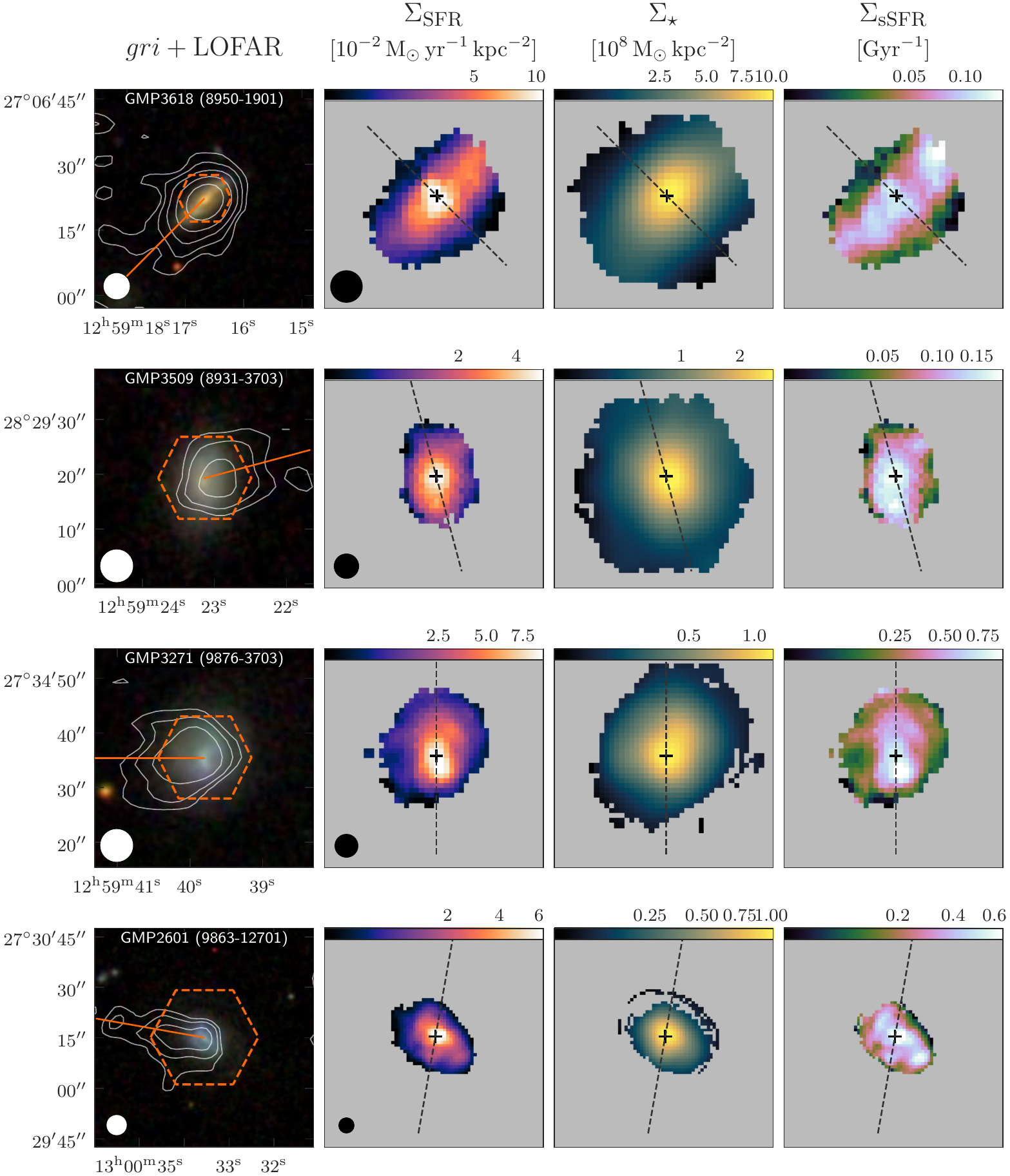}
    \caption{See Fig.~\ref{fig:panel_imgs}}
    \label{fig:panel_img2}
\end{figure*}

\begin{figure*}
    \centering
    \includegraphics[width=\textwidth]{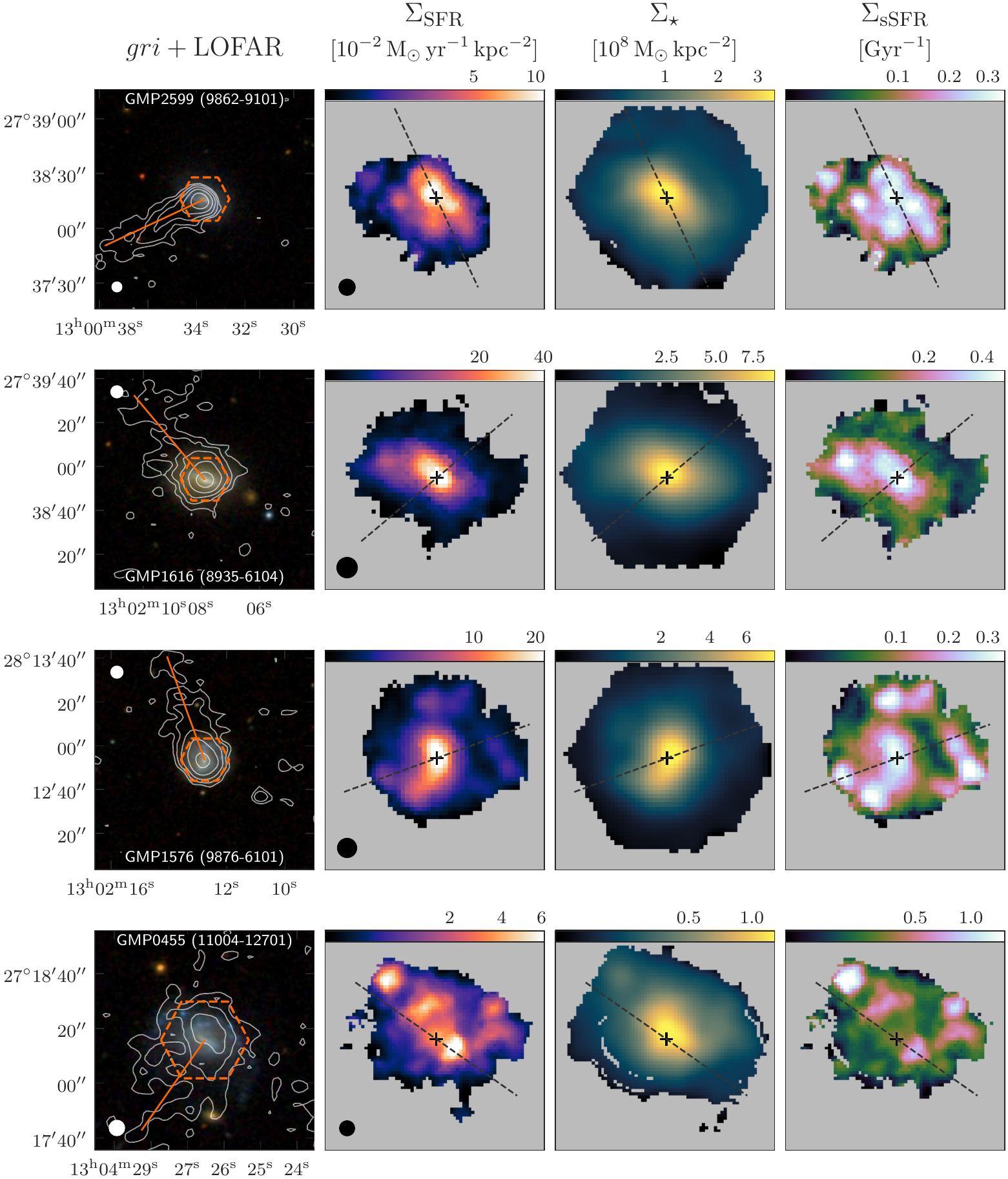}
    \caption{See Fig.~\ref{fig:panel_imgs}}
    \label{fig:panel_img3}
\end{figure*}

\begin{figure*}
    \centering
    \includegraphics[width=\textwidth]{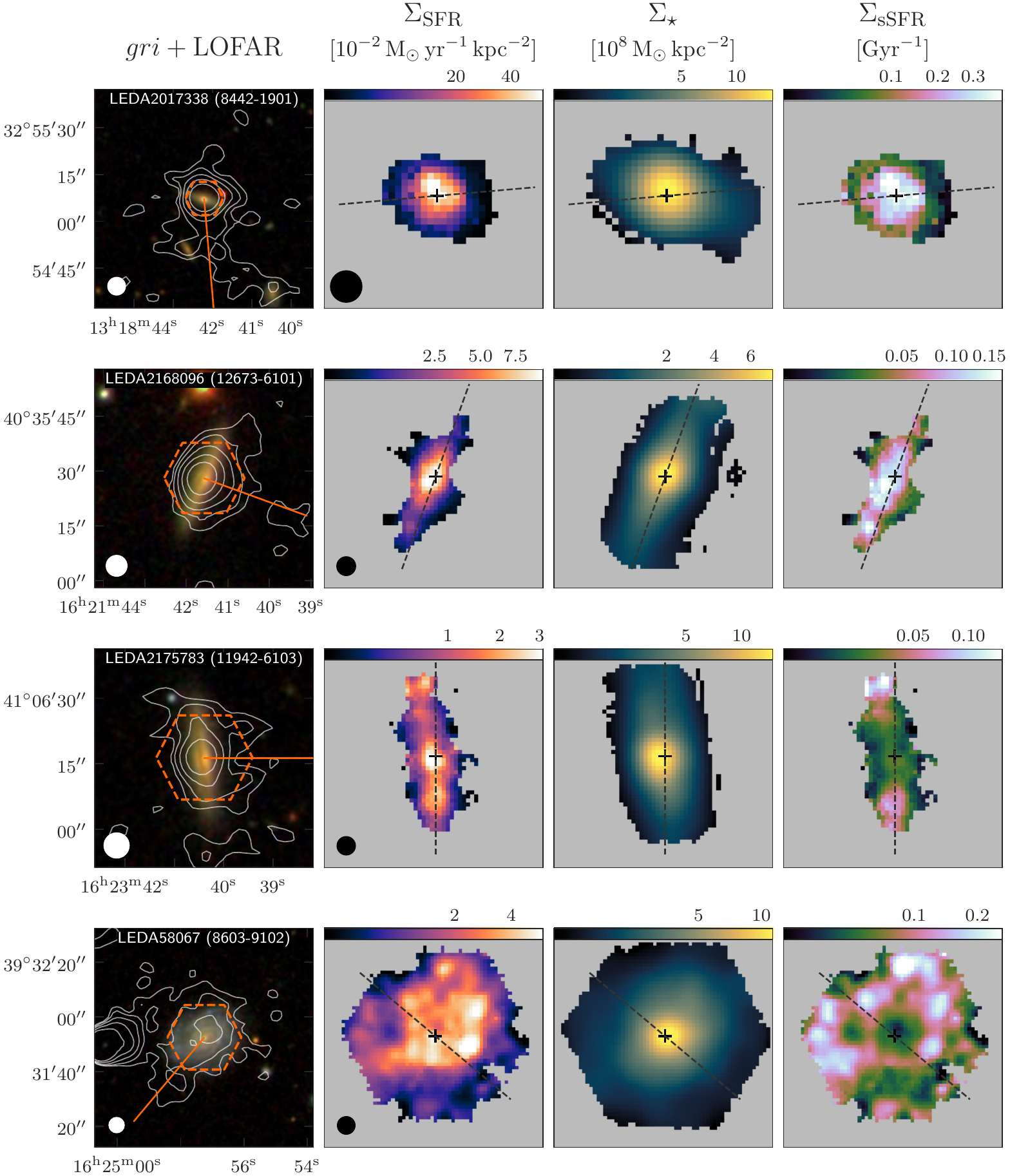}
    \caption{See Fig.~\ref{fig:panel_imgs}}
    \label{fig:panel_img4}
\end{figure*}

\begin{figure*}
    \centering
    \includegraphics[width=\textwidth]{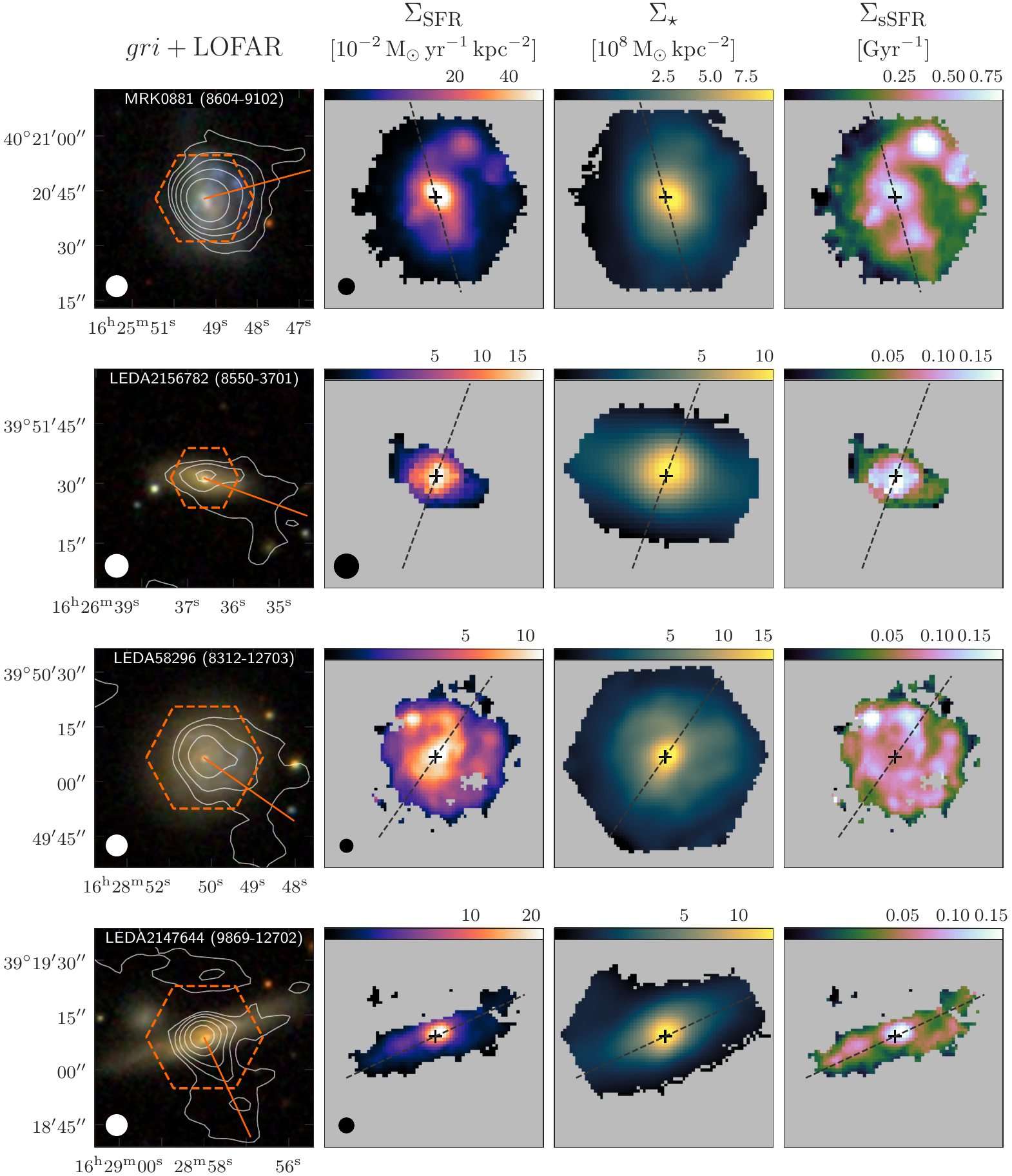}
    \caption{See Fig.~\ref{fig:panel_imgs}}
    \label{fig:panel_img5}
\end{figure*}

\begin{figure*}
    \centering
    \includegraphics[width=\textwidth]{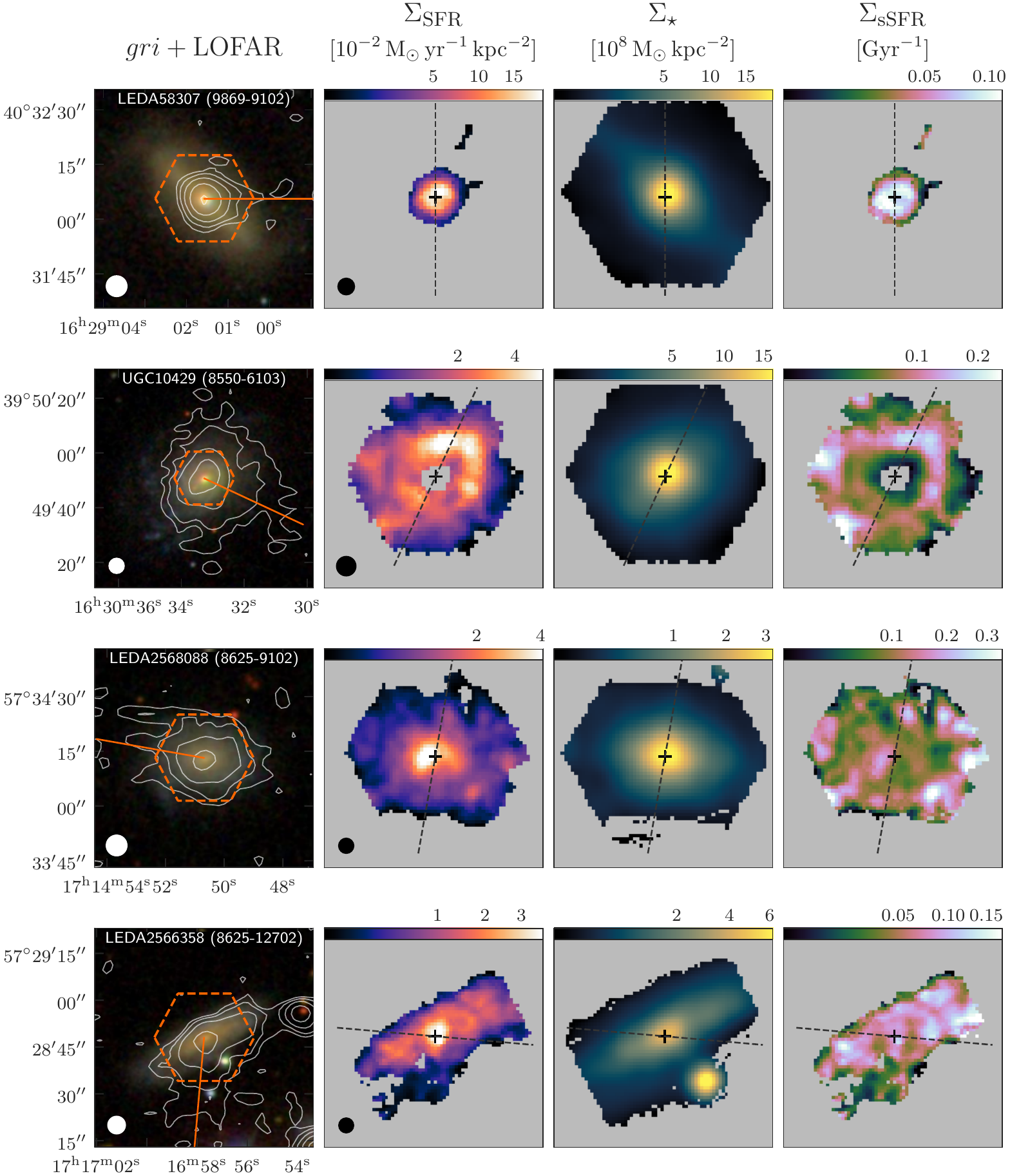}
    \caption{See Fig.~\ref{fig:panel_imgs}}
    \label{fig:panel_img6}
\end{figure*}

\begin{figure*}
    \centering
    \includegraphics[width=\textwidth]{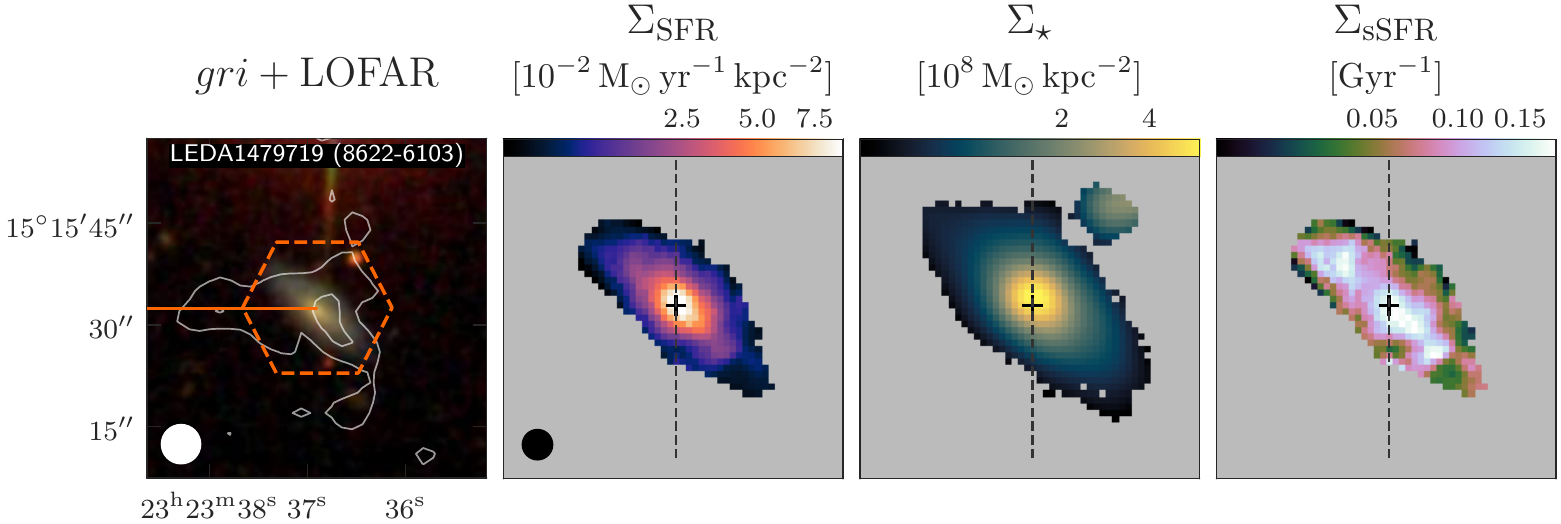}
    \caption{See Fig.~\ref{fig:panel_imgs}}
    \label{fig:panel_img7}
\end{figure*}



\end{CJK*}
\end{document}